\documentclass[12pt,iop,revtex4]{emulateapj}

\usepackage[breaklinks,colorlinks,urlcolor=blue,citecolor=blue,linkcolor=blue]{hyperref}
\usepackage{apjfonts}
\usepackage{amssymb,amsmath}
\usepackage{natbib}
\usepackage{lscape}
\usepackage{graphicx}

\newcommand{\teff}{${T}_{\mathrm{eff}}$}

\newcommand{\msun}{$M_{\odot}$}

\newcommand{\kms}{km s$^{-1}$}

\newcommand{\muhz}{$\mu$Hz}

\shorttitle{Rotation and linewidths of 27 DAVs observed by {\em Kepler}}
\shortauthors{Hermes et al.}

\begin{document}

\title{WHITE DWARF ROTATION AS A FUNCTION OF MASS AND A DICHOTOMY OF MODE LINEWIDTHS: \\ {\em KEPLER} OBSERVATIONS OF 27 PULSATING DA WHITE DWARFS THROUGH {\em K2} CAMPAIGN 8}
\author{J.~J.~Hermes\altaffilmark{1,2}, B.~T.~G\"{a}nsicke\altaffilmark{3}, Steven~D.~Kawaler\altaffilmark{4}, S.~Greiss\altaffilmark{3}, P.-E.~Tremblay\altaffilmark{3}, N.~P.~Gentile~Fusillo\altaffilmark{3}, \\ R.~Raddi\altaffilmark{3}, S.~M.~Fanale\altaffilmark{1}, Keaton~J.~Bell\altaffilmark{5}, E.~Dennihy\altaffilmark{1}, J.~T.~Fuchs\altaffilmark{1}, B.~H.~Dunlap\altaffilmark{1}, J.~C.~Clemens\altaffilmark{1}, M.~H.~Montgomery\altaffilmark{5}, D.~E.~Winget\altaffilmark{5}, P.~Chote\altaffilmark{3}, T.~R.~Marsh\altaffilmark{3}, and S.~Redfield\altaffilmark{6} }

\altaffiltext{1}{Department of Physics and Astronomy, University of North Carolina, Chapel Hill, NC 27599, USA}
\altaffiltext{2}{Hubble Fellow}
\altaffiltext{3}{Department of Physics, University of Warwick, Coventry CV4~7AL, UK}
\altaffiltext{4}{Department of Physics and Astronomy, Iowa State University, Ames, IA 50011, USA}
\altaffiltext{5}{Department of Astronomy, University of Texas at Austin, Austin, TX 78712, USA}
\altaffiltext{6}{Wesleyan University Astronomy Department, Van Vleck Observatory, 96 Foss Hill Drive, Middletown, CT 06459, USA}

\email{jjhermes@unc.edu}

\begin{abstract}

We present photometry and spectroscopy for 27 pulsating hydrogen-atmosphere white dwarfs (DAVs, a.k.a. ZZ Ceti stars) observed by the {\em Kepler} space telescope up to {\em K2} Campaign 8, an extensive compilation of observations with unprecedented duration ($>$75\,days) and duty cycle (>90\%). The space-based photometry reveals pulsation properties previously inaccessible to ground-based observations. We observe a sharp dichotomy in oscillation mode linewidths at roughly 800\,s, such that white dwarf pulsations with periods exceeding 800\,s have substantially broader mode linewidths, more reminiscent of a damped harmonic oscillator than a heat-driven pulsator. Extended {\em Kepler} coverage also permits extensive mode identification: We identify the spherical degree of 61 out of 154 unique radial orders, providing direct constraints of the rotation period for 20 of these 27 DAVs, more than doubling the number of white dwarfs with rotation periods determined via asteroseismology. We also obtain spectroscopy from 4m-class telescopes for all DAVs with {\em Kepler} photometry. Using these homogeneously analyzed spectra we estimate the overall mass of all 27 DAVs, which allows us to measure white dwarf rotation as a function of mass, constraining the endpoints of angular momentum in low- and intermediate-mass stars. We find that $0.51-0.73$\,\msun\ white dwarfs, which evolved from $1.7-3.0$\msun\ ZAMS progenitors, have a mean rotation period of 35\,hr with a standard deviation of 28\,hr, with notable exceptions for higher-mass white dwarfs. Finally, we announce an online repository for our {\em Kepler} data and follow-up spectroscopy, which we collect at {\em k2wd.org}.

\end{abstract}

\keywords{stars: white dwarfs--stars: oscillations (including pulsations)--stars: variables: general}

\section{Introduction}
\label{sec:intro}

Isolated white dwarf stars have been known to vary in brightness on short timescales for more than half a century \citep{1968ApJ...153..151L}. The roughly $2-20$\,min flux variations are known to result from surface temperature changes caused by non-radial $g$-mode pulsations, excited by surface convection (see reviews by \citealt{2008ARA&A..46..157W}, \citealt{2008PASP..120.1043F}, and \citealt{2010A&ARv..18..471A}).

Given the potential of asteroseismology to discern their interior structure, white dwarfs have been the targets of extensive observing campaigns to accurately measure their periods, which can then be compared to theoretical models to probe deep below the surface of these stellar remnants. Early observations of white dwarfs showed pulsations at many periods within the same star, further demonstrating their asteroseismic potential \citep{1972NPhS..239....2W}. However, observations from single sites were complicated by aliases arising from gaps in data collection.

Starting in the late 1980s, astronomers combated day-night aliasing by establishing the Whole Earth Telescope, a cooperative optical observing network distributed in longitude around the Earth acting as a single instrument \citep{1990ApJ...361..309N}. Dozens of astronomers trekked to remote observatories across the globe for these labor-intensive campaigns, facing complications ranging from active hurricane seasons \citep{1998ApJ...495..458O} to near-fatal stabbings \citep{2002A&A...381..122V}. Many campaigns lasted up to several weeks with duty cycles better than 70\%, delivering some of the most complete pictures of stellar structure and rotation in the 20th century for stars other than our Sun (e.g., \citealt{1991ApJ...378..326W,1994ApJ...430..839W}). 

Given the resources required, however, only a few dozen pulsating white dwarfs have been studied with the Whole Earth Telescope. Pulsating hydrogen-atmosphere white dwarfs (DAVs, a.k.a. ZZ Ceti stars) are the most numerous pulsating stars in the Galaxy, but fewer than 10 were observed with this unified network.

\begin{deluxetable*}{lcccccccccc}
\tabletypesize{\scriptsize}
\tablecolumns{11}
\tablewidth{0pc}
\tablecaption{Target selection criterion for the first 27 pulsating DA white dwarfs observed by {\em Kepler} and {\em K2}. \label{tab:selection}}
\tablehead{
        \colhead{KIC/EPIC}&
        \colhead{RA \& Dec}&
        \colhead{Alt.}&
        \colhead{$g$}&
        \colhead{($u$$-$$g$,$g$$-$$r$)}&
        \colhead{($B$$-$$R$,$R$$-$$I$)}&
        \colhead{Source}&   
        \colhead{{\em K2}}&
        \colhead{Proposal}&
        \colhead{Selection}&
        \colhead{Disc.}\\
  & (J2000.0) & Name & (mag) & (AB mag) & (AB mag) & Catalog & Field & (GO) & Method & }
\startdata
4357037 & 19 17 19.197 +39 27 19.10 & \nodata & 18.2 & (0.50, $-$0.17) & \nodata & KIS & K1 & 40109 & Colors & 1 \\
4552982 & 19 16 43.827 +39 38 49.69 & \nodata & 17.7 & \nodata & (0.02, 0.03) & SSS & K1 & 40050 & Colors & 2 \\
7594781 & 19 08 35.880 +43 16 42.36 & \nodata & 18.1 & (0.56, $-$0.16) & \nodata & KIS & K1 & 40109 & Colors & 1 \\
10132702 & 19 13 40.893 +47 09 31.28 & \nodata & 19.0 & (0.50, $-$0.16) & \nodata & KIS & K1 & 40105 & Colors & 1 \\
11911480 & 19 20 24.897 +50 17 21.32 & \nodata & 18.0 & (0.43, $-$0.16) & \nodata & KIS & K1 & 40105 & Colors & 3 \\
60017836 & 23 38 50.740 $-$07 41 19.90 & GD 1212 & 13.3 & \nodata & \nodata & \nodata & Eng & DDT & KnownZZ & 4 \\
201355934 & 11 36 04.013 $-$01 36 58.09 & WD 1133$-$013 & 17.8 & (0.46, $-$0.17) & \nodata & SDSS & C1 & 1016 & KnownZZ & 5 \\
201719578 & 11 22 21.104 +03 58 22.41 & WD 1119+042 & 18.1 & (0.39, $-$0.01) & \nodata & SDSS & C1 & 1016 & KnownZZ & 6 \\
201730811 & 11 36 55.157 +04 09 52.80 & WD 1134+044 & 17.1 & (0.49, $-$0.11) & \nodata & SDSS & C1 & 1015 & Spec. \teff & 7 \\
201802933 & 11 51 26.147 +05 25 12.90 & \nodata & 17.6 & (0.42, $-$0.19) & \nodata & SDSS & C1 & 1016 & Colors & \nodata \\
201806008 & 11 51 54.200 +05 28 39.82 & PG 1149+058 & 14.9 & (0.45, $-$0.13) & \nodata & SDSS & C1 & 1016 & KnownZZ & 8 \\
206212611 & 22 20 24.230 $-$09 33 31.09 & \nodata & 17.3 & (0.47, $-$0.14) & \nodata & SDSS & C3 & 3082 & Colors & \nodata \\
210397465 & 03 58 24.233 +13 24 30.79 & \nodata & 17.6 & (0.65, $-$0.12) & \nodata & SDSS & C4 & 4043 & Colors & \nodata \\
211596649 & 08 32 03.984 +14 29 42.37 & \nodata & 18.9 & (0.54, $-$0.18) & \nodata & SDSS & C5 & 5073 & Colors & \nodata \\
211629697 & 08 40 54.142 +14 57 08.98 & \nodata & 18.3 & (0.46, $-$0.15) & \nodata & SDSS & C5 & 5043 & Spec. \teff & \nodata \\
211914185 & 08 37 02.160 +18 56 13.38 & \nodata & 18.8 & (0.36, $-$0.21) & \nodata & SDSS & C5 & 5073 & Colors & 9 \\
211916160 & 08 56 48.334 +18 58 04.92 & \nodata & 18.9 & (0.40, $-$0.13) & \nodata & SDSS & C5 & 5073 & Colors & \nodata \\
211926430 & 09 00 41.080 +19 07 14.40 & \nodata & 17.6 & (0.47, $-$0.20) & \nodata & SDSS & C5 & 5017 & Spec. \teff & \nodata \\
228682478 & 08 40 27.839 +13 20 09.96 & \nodata & 18.2 & (0.35, $-$0.14) & \nodata & SDSS & C5 & 5043 & Colors & \nodata \\
229227292 & 13 42 11.621 $-$07 35 40.10 & \nodata & 16.6 & (0.46, $-$0.12) & \nodata & ATLAS & C6 & 6083 & Colors & \nodata \\
229228364 & 19 18 10.598 $-$26 21 05.00 & \nodata & 17.8 & \nodata & (0.10, 0.03) & SSS & C7 & 7083 & Colors & \nodata \\
220204626 & 01 11 23.888 +00 09 35.15 & WD 0108-001 & 18.4 & \nodata & \nodata & SDSS & C8 & 8018 & Spec. \teff & 7 \\
220258806 & 01 06 37.032 +01 45 03.01 & \nodata & 16.2 & (0.41, $-$0.21) & \nodata & SDSS & C8 & 8018 & Colors & \nodata \\
220347759 & 00 51 24.245 +03 39 03.79 & PHL 862 & 17.6 & (0.43, $-$0.18) & \nodata & SDSS & C8 & 8018 & Colors & \nodata \\
220453225 & 00 45 33.151 +05 44 46.96 & \nodata & 17.9 & (0.42, $-$0.15) & \nodata & SDSS & C8 & 8018 & Colors & \nodata \\
229228478 & 01 22 34.683 +00 30 25.81 & GD 842 & 16.9 & (0.43, $-$0.19) & \nodata & SDSS & C8 & 8018 & KnownZZ & 5 \\
229228480 & 01 11 00.638 +00 18 07.15 & WD 0108+000 & 18.8 & (0.45, $-$0.17) & \nodata & SDSS & C8 & 8018 & KnownZZ & 6
\enddata
\tablerefs{ Discovery of pulsations announced by (1) \citet{2016MNRAS.457.2855G}; (2) \citet{2011ApJ...741L..16H}; (3) \citet{2014MNRAS.438.3086G}; (4) \citet{2006AJ....132..831G}; (5) \citet{2010MNRAS.405.2561C}; (6) \citet{2004ApJ...607..982M}; (7) \citet{2015MNRAS.447..691P} --- search for DAVs in WD+dM systems; (8) \citet{2006AnA...450.1061V}; (9) \citet{2017ApJ...841L...2H}}
\end{deluxetable*}

The advent of space-based photometry from {\em CoRoT} and {\em Kepler} has revolutionized the study of stellar interiors, ranging from $p$-modes in solar-like oscillators on the main sequence and red-giant branch (e.g., \citealt{2013ARA&A..51..353C}) to $g$-modes in massive stars and hot subdwarfs (e.g., \citealt{2010Natur.464..259D,2011MNRAS.414.2885R}). For example, from {\em Kepler} we now have excellent constraints on core and envelope rotation in a wide range of stars \citep{2015AN....336..477A}. We have also seen new types of stochastic-like pulsations in classical heat-driven pulsators, such as {$\delta$} Scuti stars \citep{2011Natur.477..570A} and hot subdwarfs \citep{2014AA...564L..14O}.

However, this revolution did not immediately translate to white dwarf stars, few of which were identified or observed in the original {\em Kepler} mission. We aim to rectify that shortfall with an extensive search for pulsating white dwarfs observable by {\em K2}. After the failure of its second reaction wheel, {\em Kepler} has entered a new phase of observations, {\em K2}, where it observes new fields along the ecliptic every three months \citep{2014PASP..126..398H}. This has significantly expanded the number of white dwarfs available for extended observations. We have targeted all suitable candidate pulsating white dwarfs, to provide a legacy of high-cadence, extended light curves.

We present here an omnibus analysis of the first 27 DAVs observed by the {\em Kepler} space telescope, including follow-up spectroscopy to measure their atmospheric parameters. Only six of the DAVs presented here were known to pulsate prior to their {\em Kepler} observations, and only two of those six had more than 3\,hr of previous time-series photometry.

Our manuscript is organized as follows: In Section~\ref{sec:selection} we outline our target selection, and we describe our space-based photometry in Section~\ref{sec:photo}. We detail our spectroscopic observations from the 4.1-m SOAR telescope and present our full sample analysis in Section~\ref{sec:spectroscopy}. The first part of our light curve analysis, in Section~\ref{sec:hwhm}, centers on mode stability as a function of pulsation period. We summarize some of the large-scale trends we see in the DAV instablity strip as observed by {\em Kepler} and {\em K2} in Section~\ref{sec:wmp}. We then explore the asteroseismic rotation periods of 20 out of 27 of the DAVs in our sample, and place the first constraints on white dwarf rotation as a function of mass in Section~\ref{sec:rotation}. We conclude with notes on individual objects in Section~\ref{sec:individual}, as well as final discussions and conclusions in Section~\ref{sec:conclusion}. Table~\ref{tab:pulsations} in the appendix features all pulsations discovered and analyzed here: 328 entries corresponding to 154 independent modes of unique radial order ($k$) and spherical degree ($\ell$) in 27 different stars, the most extensive new listing of white dwarf pulsations ever compiled.

\section{Candidate DAV Target Selection}
\label{sec:selection}

Table~\ref{tab:selection} details the selection criteria for the first 27 DAVs observed by {\em Kepler} with short-cadence ($\sim$1\,min) exposures through {\em K2} Campaign 8 and analyzed here, including the Guest Observer (GO) program under which the target was proposed for short-cadence observations.

White dwarfs hotter than $10{,}000$\,K are most effectively selected based on their blue colors and relatively high proper motions (e.g., \citealt{2015MNRAS.448.2260G}). However, only two white dwarfs were known in the 105-square-degree {\em Kepler} field two years prior to its launch in 2009: WD\,1942+499 and WD\,1917+461. An extensive search in the year before launch --- mostly from {\em Galex} ultraviolet excess sources and faint blue objects with high proper motions --- uncovered just a dozen additional white dwarfs in the original {\em Kepler} field \citep{2010MNRAS.409.1470O}, just one pulsating \citep{2011ApJ...736L..39O}.

The most substantial deficiency for white dwarf selection came from the lack of existing blue photometry of the {\em Kepler} field, in either the SDSS-$u'$ or Johnson-$U$; in 2009 there was very little coverage from the Sloan Digital Sky Survey (SDSS) of the {\em Kepler} field. Two surveys filled this gap several years after {\em Kepler} launched: The {\em Kepler}-INT Survey (KIS), providing $U$-$g$-$r$-$i$-$H\alpha$ photometry of the entire {\em Kepler} field \citep{2012AJ....144...24G}, as well as a $U$-$V$-$B$ survey from Kitt Peak National Observatory \citep{2012PASP..124..316E}.

The KIS proved to be an excellent resource for finding new pulsating white dwarfs; we used it to discover ten new DAVs in the {\em Kepler} field \citep{2016MNRAS.457.2855G}. Unfortunately, only five were observed before the failure of the spacecraft's second reaction wheel. {\em K2} has now afforded the opportunity to observe dozens of new candidate pulsating white dwarfs in each new campaign pointing along the ecliptic.

\begin{deluxetable*}{lccccccccccccccc}
\tabletypesize{\scriptsize}
\tablecolumns{16}
\tablewidth{0pc}
\tablecaption{Details of short-cadence photometry of the first 27 pulsating DA white dwarfs observed by {\em Kepler} and {\em K2}. Columns are explained in the text. \label{tab:kepler}}
\tablehead{
        \colhead{KIC/EPIC}&
        \colhead{{\em K2}}&
        \colhead{$K_{\rm p}$}&
        \colhead{Data}&
        \colhead{CCD}&
        \colhead{Ap.}&
        \colhead{Targ.}&
        \colhead{$T_0$ (BJD$_{\rm TDB}$}&
        \colhead{Dur.}&
        \colhead{Duty}&
        \colhead{Res.}&
        \colhead{1\% FAP}&
        \colhead{0.1\% FAP}&
        \colhead{5$\langle {\rm A}\rangle$}&
        \colhead{WMP}&
        \colhead{Pub.}\\
  & Field & (mag) & Rel. & Chan. & (px) & Frac. & $-$\,2454833.0) & (d) & (\%) & (\muhz) & (ppt) & (ppt) & (ppt) & (s) &  }
\startdata
4357037 & K1 & 18.0 & 25 & 68 & 1 & 0.55 & 1488.657916 & 36.31 & 98.9 & 0.159 & 1.018 & 1.050 & 0.963 & 358.1 & \nodata \\
4552982 & K1 & 17.9 & 25 & 68 & 4 & 0.85 & 1099.398257 & 82.63 & 88.1 & 0.070 & 0.722 & 0.722 & 0.665 & 778.2 & 1 \\
7594781 & K1 & 18.2 & 25 & 45 & 1 & 0.52 & 1526.113126 & 31.84 & 99.3 & 0.182 & 0.749 & 0.781 & 0.664 & 333.3 & \nodata \\
10132702 & K1 & 18.8 & 25 & 48 & 2 & 0.61 & 1373.478019 & 97.67 & 91.8 & 0.059 & 1.431 & 1.483 & 1.441 & 749.3 & \nodata \\
11911480 & K1 & 17.6 & 25 & 2 & 2 & 0.48 & 1472.087602 & 85.86 & 84.7 & 0.067 & 0.764 & 0.794 & 0.766 & 276.9 & 2 \\
60017836 & Eng. & 13.3 & $-$1 & 76 & 93 & 0.97 & 1860.040459 & 8.91 & 98.9 & 0.649 & 0.284 & 0.297 & 0.172 & 1019.1 & 3 \\
201355934 & C1 & 17.9 & 14 & 39 & 4 & 0.33 & 1975.168594 & 77.52 & 92.9 & 0.075 & 0.478 & 0.493 & 0.462 & 208.2 & \nodata \\
201719578 & C1 & 18.1 & 14 & 67 & 11 & 0.88 & 1975.168222 & 78.80 & 94.7 & 0.073 & 0.956 & 0.993 & 0.883 & 740.9 & \nodata \\
201730811 & C1 & 17.1 & 14 & 46 & 8 & 0.98 & 1975.168490 & 77.18 & 94.7 & 0.075 & 0.253 & 0.262 & 0.242 & 248.4 & 4 \\
201802933 & C1 & 17.7 & 14 & 29 & 11 & 0.95 & 1975.168648 & 77.24 & 94.9 & 0.075 & 0.605 & 0.630 & 0.592 & 266.2 & \nodata \\
201806008 & C1 & 15.0 & 14 & 29 & 33 & 0.91 & 1975.168653 & 78.91 & 95.3 & 0.073 & 0.311 & 0.322 & 0.162 & 1030.4 & \nodata \\
206212611 & C3 & 17.4 & 10 & 46 & 6 & 0.97 & 2144.093232 & 69.18 & 98.3 & 0.084 & 0.304 & 0.314 & 0.321 & 1204.4 & \nodata \\
210397465 & C4 & 17.7 & 10 & 54 & 8 & 0.96 & 2228.790357 & 70.90 & 98.7 & 0.082 & 0.945 & 0.980 & 0.879 & 841.0 & \nodata \\
211596649 & C5 & 19.0 & 10 & 58 & 4 & 0.99 & 2306.600768 & 74.84 & 98.7 & 0.077 & 1.695 & 1.762 & 1.809 & 288.7 & \nodata \\
211629697 & C5 & 18.4 & 10 & 38 & 5 & 0.98 & 2306.600916 & 74.84 & 98.7 & 0.077 & 0.758 & 0.784 & 0.772 & 1045.2 & \nodata \\
211914185 & C5 & 18.9 & 10 & 46 & 5 & 0.97 & 2306.600805 & 74.84 & 98.7 & 0.077 & 0.922 & 0.955 & 0.976 & 161.7 & 5 \\
211916160 & C5 & 19.0 & 10 & 25 & 6 & 0.82 & 2306.601137 & 74.84 & 98.5 & 0.077 & 1.191 & 1.237 & 1.256 & 201.3 & \nodata \\
211926430 & C5 & 17.7 & 10 & 9 & 6 & 0.96 & 2306.601189 & 74.84 & 98.4 & 0.077 & 0.471 & 0.486 & 0.456 & 244.2 & \nodata \\
228682478 & C5 & 18.3 & 10 & 39 & 10 & 0.74 & 2306.600932 & 74.84 & 98.2 & 0.077 & 0.395 & 0.410 & 0.413 & 301.1 & \nodata \\
229227292 & C6 & 16.7 & 8 & 48 & 4 & 0.98 & 2384.453493 & 78.93 & 98.2 & 0.073 & 0.347 & 0.357 & 0.277 & 964.5 & \nodata \\
229228364 & C7 & 17.9 & 9 & 39 & 7 & 0.98 & 2467.250743 & 78.72 & 98.0 & 0.074 & 0.487 & 0.508 & 0.495 & 1116.4 & \nodata \\
220204626 & C8 & 17.2 & 11 & 54 & 19 & 0.83 & 2559.058649 & 78.72 & 98.3 & 0.074 & 0.805 & 0.832 & 0.846 & 642.7 & \nodata \\
220258806 & C8 & 16.4 & 11 & 59 & 8 & 0.90 & 2559.058611 & 78.72 & 96.0 & 0.074 & 0.155 & 0.161 & 0.154 & 206.0 & \nodata \\
220347759 & C8 & 17.7 & 11 & 64 & 8 & 0.83 & 2559.058460 & 78.72 & 96.5 & 0.074 & 0.458 & 0.470 & 0.476 & 209.3 & \nodata \\
220453225 & C8 & 18.0 & 11 & 68 & 9 & 0.93 & 2559.058403 & 78.72 & 96.7 & 0.074 & 0.625 & 0.646 & 0.631 & 1034.9 & \nodata \\
229228478 & C8 & 17.0 & 11 & 35 & 22 & 0.97 & 2559.058796 & 78.72 & 98.5 & 0.074 & 0.514 & 0.534 & 0.535 & 141.8 & \nodata \\
229228480 & C8 & 18.9 & 11 & 54 & 8 & 0.89 & 2559.058649 & 78.72 & 98.6 & 0.074 & 2.501 & 2.586 & 2.618 & 280.0 & \nodata
\enddata
\tablerefs{ {\em Kepler}/{\em K2} data analyzed by (1) \citet{2015ApJ...809...14B}; (2) \citet{2014MNRAS.438.3086G}; (3) \citet{2014ApJ...789...85H}; (4) \citet{2015MNRAS.451.1701H}; (5) \citet{2017ApJ...841L...2H}}
\end{deluxetable*}

DAV pulsations, excited by a hydrogen partial-ionization zone, occur in a narrow temperature range between roughly $12{,}600-10{,}600$\,K for canonical-mass (0.6\,\msun) white dwarfs \citep{2015ApJ...809..148T}. Therefore, the most efficient way to select candidate DAVs is to look for those with effective temperatures within the DAV instability strip. Empirically, this is best accomplished from Balmer-line fits to low-resolution spectroscopy, which is available for many white dwarfs in SDSS (e.g., \citealt{2004ApJ...607..982M}).

We targeted four new DAVs observed through {\em K2} Campaign 8 based on serendipitous SDSS spectroscopy, which revealed atmospheric parameters within the empirical instability strip \citep{2013ApJS..204....5K}. The majority of our new DAV candidates did not have spectroscopy and were selected based on their ($u$$-$$g$,$g$$-$$r$) colors, mostly from SDSS \citep{2015MNRAS.448.2260G}, plus one from ($u$$-$$g$,$g$$-$$r$) colors from an early data release of the VST ATLAS survey \citep{2017MNRAS.469..621G}. Three color-selected white dwarfs in Campaign~5 were proposed for short-cadence {\em K2} observations for possible planetary transits (GO program 5073).

We also selected two DAVs from the Supercosmos Sky Survey (SSS) catalog of \citet{2011MNRAS.417...93R}, identified as white dwarfs based on their proper motions and selected as DAV candidates by their ($B$$-$$R$, $R$$-$$I$) colors. This includes the white dwarf with the longest space-based light curve, observed in the original {\em Kepler} field \citep{2011ApJ...741L..16H}.

Finally, six DAVs observed by {\em K2} were known to pulsate before the launch of the spacecraft (KnownZZ). Our full selection information is summarized in Table~\ref{tab:selection}, including the publication announcing the discovery of variability when relevant. We plan to publish our full list of candidate DAVs, including limits on those not observed to vary with {\em K2} observations, at the end of the mission.

We note that in fact 29 DAVs have been observed by the {\em Kepler} spacecraft through Campaign 8. However, we exclude two DAVs from our analysis here because they each have more than 100 significant periodicities, deserving of their own individual analyses. Neither star appears to be in tension with our general trends or results. The excluded DAVs (EPIC\,211494257 and EPIC\,212395381) will be detailed in two forthcoming publications.

\section{Space-Based {\em Kepler} Photometry}
\label{sec:photo}

All observations analyzed here were collected by the {\em Kepler} spacecraft with short-cadence exposures, which are co-adds of $9\times6.02$\,s exposures for a total exposure time of 58.85\,s, including readout overheads \citep{2010ApJ...713L.160G}. Full details of the raw and processed {\em Kepler} and {\em K2} observations are summarized in Table~\ref{tab:kepler}.

For the five DAVs observed in the original {\em Kepler} mission, we analyzed light curves processed by the GO office \citep{2012PASP..124..985S,2012PASP..124.1000S}. We produced final light curves using the calculated {\sc PDCSAP} flux and flux uncertainties, after iteratively clipping all points 5$\sigma$ from the light curve median and fitting out a second-order polynomial to correct for any long-term instrumental drift. Only two DAVs in the original mission were observed for more than one quarter, and their full light curves have been analyzed elsewhere (\citealt{2015ApJ...809...14B} for KIC\,4552982 and \citealt{2014MNRAS.438.3086G} for KIC\,11911480). Here we analyze only Q12 for KIC\,4552982 and Q16 for KIC\,11911480, to remain relatively consistent with the duration of the other DAVs observed with {\em K2}. All original-mission observations were extracted from data release 25, which includes updated smear corrections for short-cadence data\footnote{\href{https://keplerscience.arc.nasa.gov/data/documentation/KSCI-19080-002.pdf}{https://keplerscience.arc.nasa.gov/data/documentation/KSCI-19080-002.pdf}}, which is the cause of the amplitude discrepancy in KIC\,11911480 observed by \citet{2014MNRAS.438.3086G}.

Reducing data obtained during the two-reaction-wheel-controlled {\em K2} mission requires more finesse, since the spacecraft checks its roll orientation roughly every six hours and, if necessary, fires its thrusters to remain accurately pointed. Thruster firings introduce discontinuities into the photometry.

For each DAV observed with {\em K2}, we have downloaded its short-cadence Target Pixel File from Mikulski Archive for Space Telescopes and processed it using the {\sc PyKE} software package managed by the {\em Kepler} GO office \citep{2012ascl.soft08004S}. We began by choosing a fixed aperture that maximized the signal-to-noise (S/N) of our target, extracted the target flux, and subsequently fit out a second-order polynomial to 3-day segments to flatten longer-term instrumental trends. We subsequently used the {\sc KEPSFF} task by \citet{2014PASP..126..948V} to mitigate the {\em K2} motion-correlated systematics. Finally, we iteratively clipped all points 5$\sigma$ from the light curve median and fit out a second-order polynomial to the whole dataset to produce a final light curve\footnote{All reduced light curves are available online at {\em \href{http://www.k2wd.org}{http://www.k2wd.org}.}}.

We list the size of the final fixed aperture (Ap.) used for our extractions in Table~\ref{tab:kepler}, along with the CCD channel from which the observations were read out. Channel 48 and especially Channel 58 are susceptible to rolling band pattern noise, which can lead to long-term systematics in the light curve (see Section 3 of \citealt{2017MNRAS.468.1946H} and references therein).

\begin{figure}
\centering{\includegraphics[width=0.995\columnwidth]{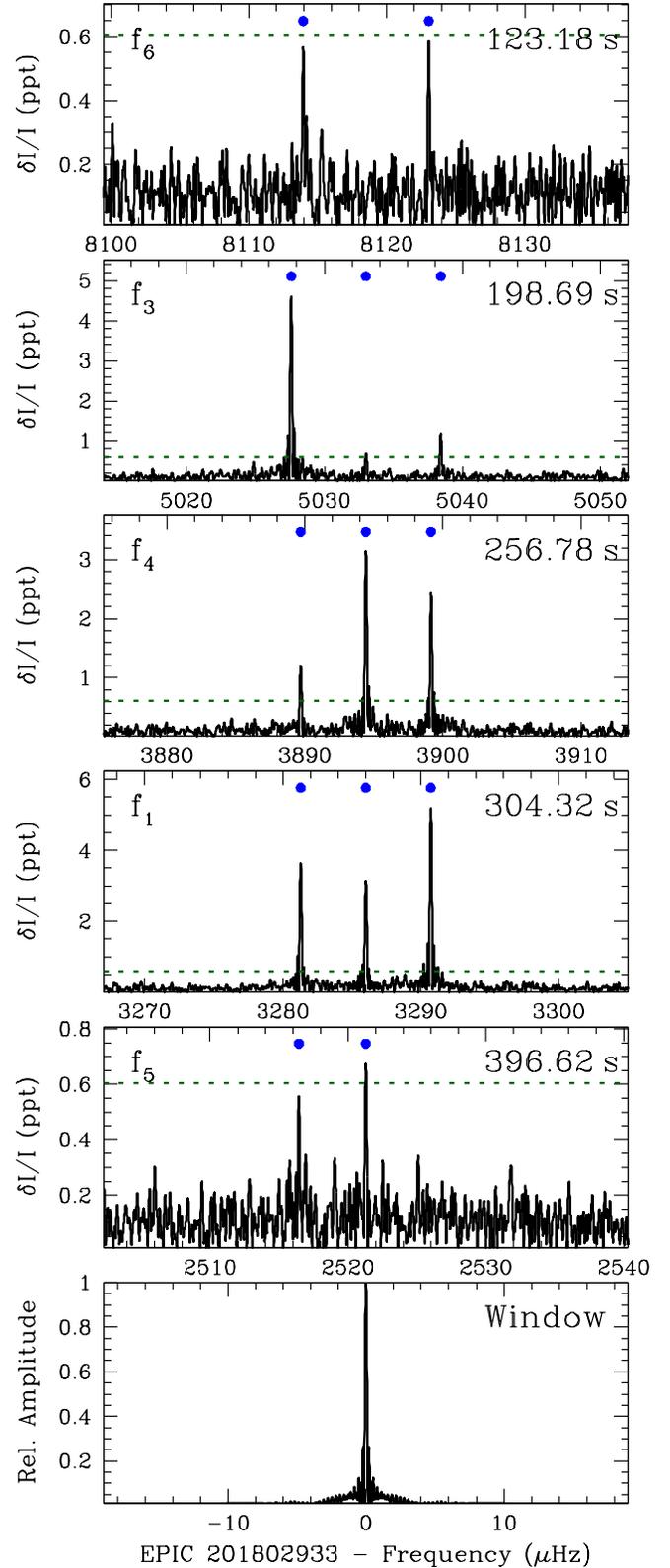}}
\caption{A detailed look at the Fourier transform of EPIC 201802933 (SDSSJ1151+0525, $K_{\rm p}$=17.7\,mag), a new DAV discovered with {\em K2} and representative of most of the hot DAVs analyzed here. The green dotted line shows the 1\% FAP (see text). The {\em K2} photometry features an exceptionally sharp spectral window (shown in the bottom panel), simplifying mode identification. All of the pulsations shown here are most simply interpreted as components of five different radial orders ($k$) of $\ell=1$ dipole modes. The median frequency splitting of the five dipole modes is $\delta f~=~4.7$\,\muhz, which corresponds to a roughly 1.3-day rotation period (see Section~\ref{sec:rotation}).
 \label{fig:ftc1}}
\end{figure}

We note the ``Targ. Frac.'' column of Table~\ref{tab:kepler}, which estimates (on a scale from $0.0-1.0$) the total fraction of the flux contained in the aperture that belongs to the target. This is important for crowded fields, since each {\em Kepler} pixel covers 4\arcsec\ on a side. Amplitudes have not been adjusted by this factor, which should be applied when comparing pulsation amplitudes reported here to ground-based amplitudes or star-to-star amplitude differences. Our target fraction values for DAVs in the original mission come from the {\em Kepler} Input Catalog \citep{2011AJ....142..112B}. {\em K2} targets do not have such estimates, so we have calculated the target fraction for DAVs in {\em K2} by modeling the point spread function of our target and comparing the modeled flux of the target to the total flux observed in the extracted aperture. We perform this calculation using the pixel-response function tool {\sc KEPPRF} in {\sc PyKE}.

The light curves described here are the longest systematic look ever undertaken of white dwarf pulsations. All but three of the 27 DAVs have more than 69\,days of observations with high duty cycles, in most cases exceeding 98\%. This long-baseline coverage provides frequency resolution to better than 0.08\,\muhz\ in almost all cases. The high duty cycle ensures an exceptionally clean spectral window, easing the interpretation of the frequency spectra and simplifying mode identification.

Figure~\ref{fig:ftc1} highlights a representative amplitude spectrum for a hot DAV: EPIC\,201802933 (SDSSJ1151+0525, $K_{\rm p}$=17.7\,mag), observed for more than 77\,days in {\em K2} Campaign 1. All Fourier transforms we compute have been oversampled by a factor of 20, calculated from the software package {\sc Period04} \citep{2005CoAst.146...53L}. We fit for and removed all instrumental artifacts arising from the long-cadence sampling rate, at integer multiples of roughly 566.48\,\muhz\ \citep{2010ApJ...713L.160G}. All panels of Figure~\ref{fig:ftc1} are on the same frequency scale, including the spectral window. 

We computed a significance threshold for all DAVs from a shuffled simulation of the data, as described in \citet{2015MNRAS.451.1701H}. In summary, we keep the time sampling of our observed light curves but randomly shuffle the flux values to create $10{,}000$ synthetic light curves, noting the highest peak in the Fourier transform for each synthetic dataset. We set our 1\% (0.1\%) False Alarm Probability (FAP) at the value for which 99\% (99.9\%) of these synthetic Fourier transforms do not have a peak exceeding that amplitude. We also list in Table~\ref{tab:kepler} the value for five times the average amplitude of the entire Fourier transform, 5$\langle {\rm A}\rangle$.

\begin{deluxetable*}{lccccccccccccc}
\tabletypesize{\scriptsize}
\tablecolumns{14}
\tablewidth{0pc}
\tablecaption{Follow-up spectroscopy of the first 27 pulsating DA white dwarfs observed by {\em Kepler} and {\em K2}. \label{tab:spectroscopy}}
\tablehead{
        \colhead{KIC/EPIC}&
        \colhead{$g$}&
        \colhead{Facility, Night of}&
        \colhead{Exp.}&
        \colhead{Seeing}&
        \colhead{Airm.}&
        \colhead{S/N}&
        \colhead{${T}_{\mathrm{eff}}$-1D}&
        \colhead{$\Delta{T}_{\mathrm{eff}}$}&
        \colhead{$\log{g}$-1D}&
        \colhead{$\Delta\log{g}$}&
        \colhead{${T}_{\mathrm{eff}}$-3D}&
        \colhead{$\log{g}$-3D}&
        \colhead{$M_{\mathrm{WD}}$}\\
  & (mag) &  & (s) & ($\arcsec$) & Avg. &  & (K) & (K) & (cgs) & (cgs) & (K) & (cgs) & (\msun) }
\startdata
4357037 & 18.2 & WHT, 2013 Jun 06 & 3$\times$1200 & 0.9 & 1.07 & 77 & 12750 & 240 & 8.020 & 0.062 & 12650 & 8.013 & 0.62 \\
4552982$^{\dagger}$ & 17.7 & WHT, 2014 July 25 & 5$\times$1200 & 0.7 & 1.06 & 76 & 11240 & 160 & 8.280 & 0.053 & 10950 & 8.113 & 0.67 \\
7594781 & 18.1 & WHT, 2014 July 25 & 6$\times$1800 & 0.8 & 1.11 & 70 & 12040 & 210 & 8.170 & 0.059 & 11730 & 8.107 & 0.67 \\
10132702 & 19.0 & WHT, 2013 Jun 06 & 3$\times$1500 & 0.9 & 1.06 & 70 & 12220 & 240 & 8.170 & 0.072 & 11940 & 8.123 & 0.68 \\
11911480 & 18.0 & WHT, 2013 Jun 07 & 3$\times$1200 & 1.0 & 1.22 & 68 & 11880 & 190 & 8.020 & 0.061 & 11580 & 7.959 & 0.58 \\
60017836 & 13.3 & SOAR, 2016 Jul 14 & 5$\times$30 & 2.1 & 1.11 & 210 & 11280 & 140 & 8.144 & 0.040 & 10980 & 7.995 & 0.60 \\
201355934 & 17.8 & SOAR, 2017 Apr 20 & 5$\times$300 & 1.2 & 1.13 & 73 & 12050 & 170 & 8.016 & 0.047 & 11770 & 7.972 & 0.59 \\
201719578 & 18.1 & SOAR, 2017 Apr 14 & 8$\times$300 & 1.4 & 1.22 & 62 & 11390 & 150 & 8.068 & 0.047 & 11070 & 7.941 & 0.57 \\
201730811 & 17.1 & SOAR, 2015 Jan 28 & 4$\times$420 & 1.0 & 1.24 & 139 & 12600 & 170 & 7.964 & 0.043 & 12480 & 7.956 & 0.58 \\
201802933 & 17.6 & SOAR, 2016 Feb 14 & 3$\times$360 & 1.4 & 1.23 & 75 & 12530 & 180 & 8.136 & 0.046 & 12330 & 8.114 & 0.68 \\
201806008$^{\dagger}$ & 14.9 & SOAR, 2016 Jun 22 & 7$\times$180 & 1.1 & 1.29 & 161 & 11200 & 140 & 8.182 & 0.040 & 10910 & 8.019 & 0.61 \\
206212611 & 17.3 & SOAR, 2016 Jul 14 & 3$\times$300 & 1.4 & 1.07 & 77 & 11120 & 150 & 8.170 & 0.048 & 10830 & 7.999 & 0.60 \\
210397465 & 17.6 & SOAR, 2016 Sep 01 & 5$\times$300 & 1.8 & 1.27 & 50 & 11520 & 160 & 7.782 & 0.054 & 11200 & 7.713 & 0.45 \\
211596649 & 18.9 & SOAR, 2017 Feb 28 & 7$\times$600 & 1.8 & 1.22 & 51 & 11890 & 180 & 7.967 & 0.057 & 11600 & 7.913 & 0.56 \\
211629697$^{\dagger}$ & 18.3 & SOAR, 2016 Dec 27 & 13$\times$420 & 2.6 & 1.29 & 71 & 10890 & 150 & 7.950 & 0.060 & 10600 & 7.772 & 0.48 \\
211914185 & 18.8 & SOAR, 2017 Jan 25 & 9$\times$600 & 2.0 & 1.38 & 61 & 13620 & 380 & 8.437 & 0.058 & 13590 & 8.434 & 0.88 \\
211916160 & 18.9 & SOAR, 2017 Apr 21 & 10$\times$480 & 1.2 & 1.60 & 56 & 11820 & 170 & 8.027 & 0.053 & 11510 & 7.958 & 0.58 \\
211926430 & 17.6 & SOAR, 2016 Dec 27 & 11$\times$420 & 2.2 & 1.30 & 98 & 11740 & 160 & 8.065 & 0.045 & 11420 & 7.982 & 0.59 \\
228682478 & 18.2 & SOAR, 2016 Jan 08 & 4$\times$600 & 1.3 & 1.28 & 99 & 12340 & 170 & 8.226 & 0.046 & 12070 & 8.184 & 0.72 \\
229227292$^{\dagger}$ & 16.6 & SOAR, 2016 Feb 14 & 6$\times$180 & 1.2 & 1.20 & 103 & 11530 & 150 & 8.146 & 0.042 & 11210 & 8.028 & 0.62 \\
229228364$^{\dagger}$ & 17.8 & SOAR, 2016 Jul 05 & 8$\times$420 & 1.9 & 1.06 & 147 & 11330 & 140 & 8.172 & 0.043 & 11030 & 8.026 & 0.62 \\
220204626 & 18.4 & SOAR, 2016 Jul 05 & 7$\times$420 & 1.7 & 1.22 & 74 & 11940 & 250 & 8.255 & 0.061 & 11620 & 8.173 & 0.71 \\
220258806 & 16.2 & SOAR, 2016 Jul 14 & 3$\times$180 & 2.1 & 1.18 & 134 & 12890 & 200 & 8.093 & 0.047 & 12800 & 8.086 & 0.66 \\
220347759 & 17.6 & SOAR, 2016 Jul 05 & 5$\times$300 & 1.5 & 1.32 & 81 & 12860 & 200 & 8.087 & 0.047 & 12770 & 8.080 & 0.66 \\
220453225$^{\dagger}$ & 17.9 & SOAR, 2016 Jul 05 & 7$\times$300 & 1.6 & 1.49 & 72 & 11540 & 160 & 8.153 & 0.045 & 11220 & 8.035 & 0.62 \\
229228478 & 16.9 & SOAR, 2014 Oct 13 & 6$\times$180 & 1.3 & 1.19 & 75 & 12610 & 180 & 7.935 & 0.045 & 12500 & 7.929 & 0.57 \\
229228480 & 18.8 & SOAR, 2016 Aug 07 & 7$\times$600 & 1.0 & 1.17 & 65 & 12640 & 190 & 8.202 & 0.047 & 12450 & 8.181 & 0.72
\enddata
\tablerefs{ We use a $^{\dagger}$ symbol to mark the first six outbursting white dwarfs \citep{2015ApJ...809...14B,2015ApJ...810L...5H,2016ApJ...829...82B,2017ASPC..509..303B}}
\end{deluxetable*}

For all DAVs we adopt the 1\% FAP as our significance threshold and produce a period list of observed pulsations, detailed in Table~\ref{tab:pulsations} in the appendix. Our first set of uncertainties on the period, frequency, amplitude, and phase in Table~\ref{tab:pulsations} arise from a simultaneous non-linear least-square fit to the significant peaks, calculated with {\sc Period04}. All light curves are barycentric corrected, and the phases in Table~\ref{tab:pulsations} are relative to the first mid-exposure time ($T_0$) listed in in Table~\ref{tab:kepler}. The frequencies of variability are not in the stellar rest frame, but many have high enough precision that they should eventually be corrected for the Doppler and gravitational redshift of the white dwarf (e.g., \citealt{2014MNRAS.445L..94D}).

Importantly, we are able to identify the spherical degree ($\ell$) of 61 of the 154 independent modes in these DAVs (roughly 40\%), based on common frequency patterns in the Fourier transforms, which in turn illuminates the rotation periods of these stellar remnants (see Section~\ref{sec:rotation}). When identified, we group in Table~\ref{tab:pulsations} each set of frequencies of the same radial order ($k$) and spherical degree ($\ell$), and include the measured frequency splittings. We relax the significance threshold for a handful of modes that share common frequency splittings, which occasionally helps with identifying the correct azimuthal order ($m$) for modes present. We have not attempted to quantify the exact radial orders of any of the pulsations here, but leave that to future asteroseismic analysis of these stars.

\section{Follow-Up WHT \& SOAR Spectroscopy}
\label{sec:spectroscopy}

We complemented our space-based photometry of these 27 DAVs by determining their atmospheric parameters based on model-atmosphere fits to follow-up spectroscopy obtained from two 4-m-class, ground-based facilities. We detail these spectroscopic observations and their resultant fits in Table~\ref{tab:spectroscopy}. Our spectra have been obtained, reduced, and fit in a homogenous way to minimize systematics.

DAVs in the original {\em Kepler} mission field are at such high declination ($\delta>+39$\,deg) that they can only be observed from the Northern hemisphere, so we used the Intermediate-dispersion Spectrograph and Imaging System (ISIS) instrument mounted on the 4.2-m William Herschel Telescope (WHT) on the island of La Palma for these five northern DAVs. Their spectra were taken with a 600 line mm$^{-1}$ grating and cover roughly $3800-5100$\,\AA\ at roughly $2.0$\,\AA\ resolution using the blue arm of ISIS. A complete journal of observations for these five DAVs is detailed in \citet{2016MNRAS.457.2855G}, and we only list in Table~\ref{tab:spectroscopy} the night for which most observations were obtained.

\begin{figure*}[t]
\centering{\includegraphics[width=0.72\textwidth]{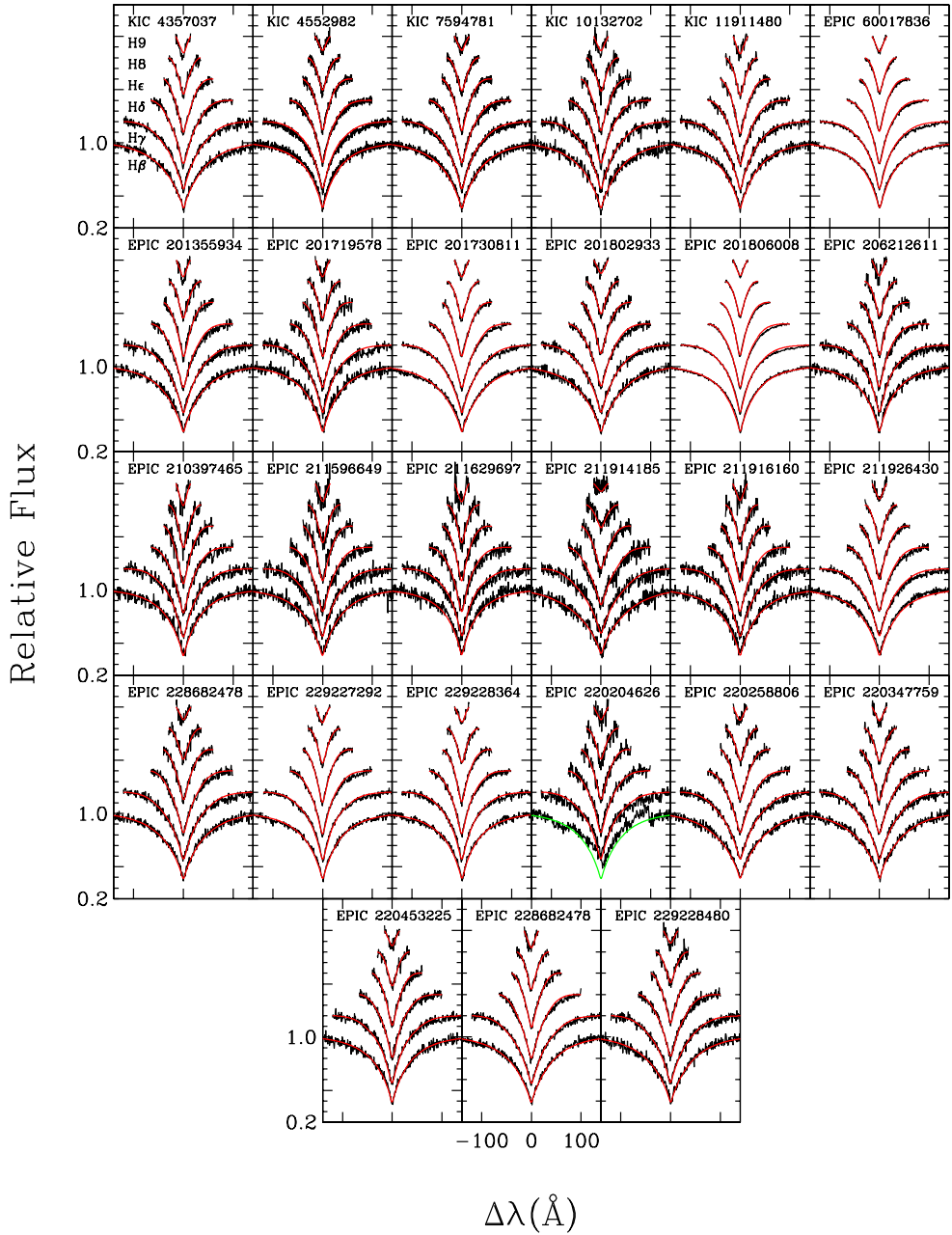}}
\caption{The averaged, normalized spectra for the 27 DAVs observed through {\em K2} Campaign 8, overplotted with the best-fit atmospheric parameters detailed in Table~\ref{tab:spectroscopy}. We observed the five DAVs from the original {\em Kepler} mission with the ISIS spectrograph on the 4.2-m William Herschel Telescope; all others were observed with the Goodman spectrograph on the 4.1-m SOAR telescope. \citet{2011ApJ...730..128T} describes the models and fitting procedures, which use ML2/$\alpha=0.8$; atmospheric parameters have been corrected for the three-dimensional dependence of convection \citep{2013A&A...559A.104T}. \label{fig:balmerfits}}
\end{figure*}

\begin{figure*}[t]
\centering{\includegraphics[width=0.995\textwidth]{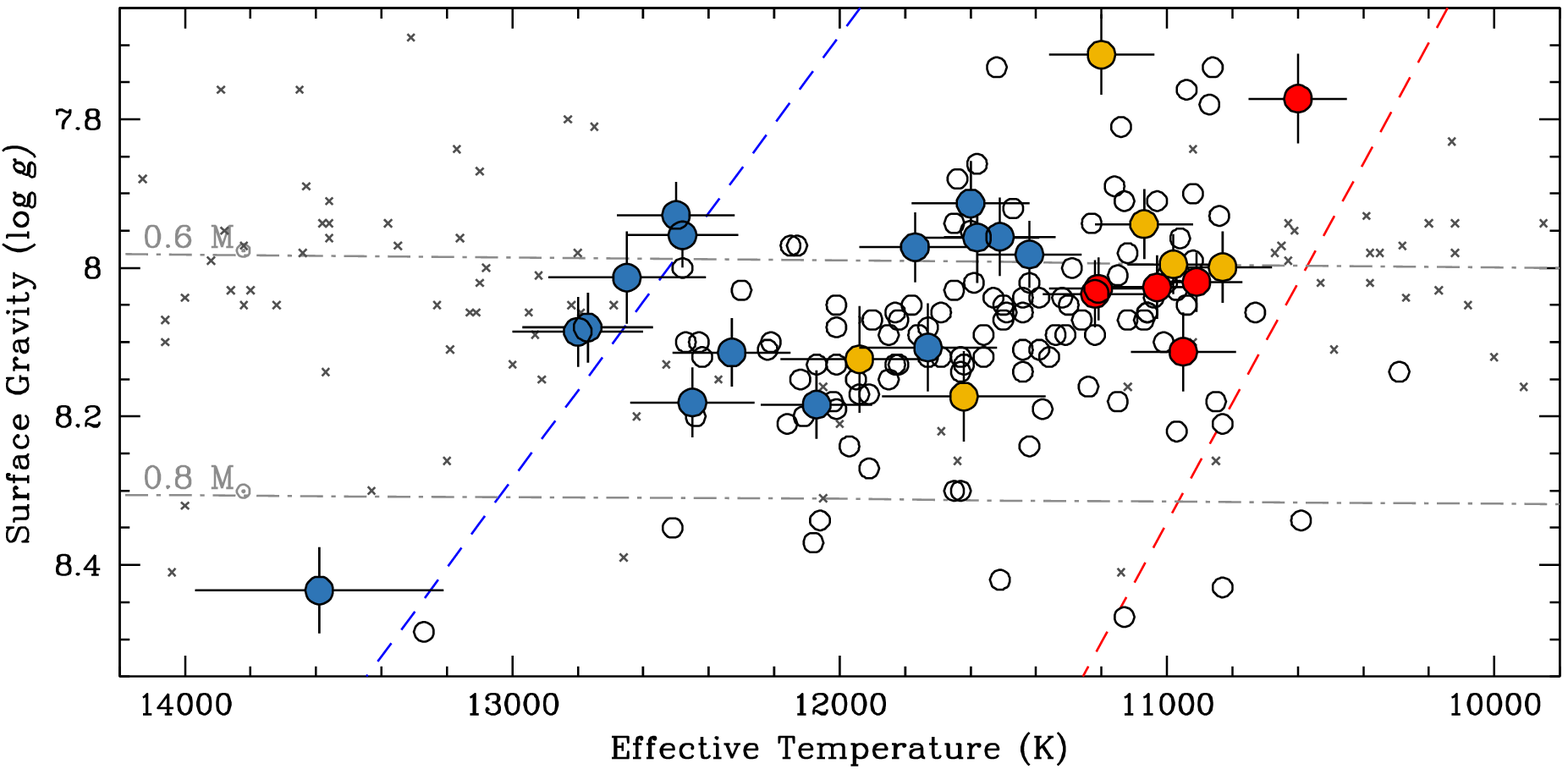}}
\caption{The DAVs with {\em Kepler} data in the context of the empirical DAV instability strip, demarcated with blue and red dashed lines most recently updated by \citet{2015ApJ...809..148T}. We mark all DAVs with weighted mean pulsation periods shorter than 600\,s in blue, and those greater than 600\,s in gold (see Section~\ref{sec:wmp}). Cool DAVs which show outbursts are marked in red (see Section~\ref{sec:wmp}). Known DAVs from ground-based observations are shown with open circles, and DAs not observed to vary to higher than 4 ppt are marked with grey crosses \citep{2011ApJ...743..138G,2011ApJ...730..128T}. All atmospheric parameters have been analyzed with the same models, using ML2/$\alpha$=0.8, and corrected for the 3D-dependence of convection \citep{2013A&A...559A.104T}. Dash-dotted gray lines show cooling tracks for 0.6\,\msun\ and 0.8\,\msun\ white dwarfs \citep{2001PASP..113..409F}. \label{fig:istrip}}
\end{figure*}

All {\em K2} fields lie along the ecliptic, so we obtained spectroscopy of all {\em K2} DAVs with the Goodman spectrograph \citep{2004SPIE.5492..331C} mounted on the 4.1-m Southern Astrophysical Research (SOAR) telescope on Cerro Pach\'{o}n in Chile. Using a high-throughput 930 line mm$^{-1}$ grating, we adopted grating and camera angles (13 degrees and 24 degrees, respectively) which yield wavelength coverage from roughly $3600-5200$\,\AA. With 2$\times$2 binning, our dispersion is 0.84\,\AA\ pixel$^{-1}$. To capture as much flux as possible, we use a 3\arcsec\ slit, so our spectral resolution is seeing limited, roughly 3\,\AA\ in 1.0\arcsec\ seeing and roughly 4\,\AA\ in 1.4\arcsec\ seeing.

For each DAV, we obtained a series of consecutive spectra covering at least two cycles of the highest-amplitude pulsation, as derived from the {\em Kepler} photometry. We also made attempts to observe our targets at minimum airmass, which we list in Table~\ref{tab:spectroscopy}, along with the mean seeing during the observations, measured from the mean FWHM in the spatial direction of the two-dimensional spectra.

All our spectra were debiased and flat-fielded using a quartz lamp, using standard {\sc STARLINK} routines \citep{2014ASPC..485..391C}. They were optimally extracted \citep{1986PASP...98..609H} using the software {\sc PAMELA}. We then used {\sc MOLLY} \citep{1989PASP..101.1032M} to wavelength calibrate and apply a heliocentric correction, and perform a final weighted average of the one-dimensional (1D) spectra. We flux calibrated each spectrum to a spectrophotometric standard observed using the same setup at a similar airmass, and used a scalar to normalize this final spectrum to the $g$ magnitude from either the SDSS, VST/ATLAS, APASS, or KIS photometric surveys, in order to correct its final absolute flux calibration.

We use the mean seeing to compute the spectral resolution, which informs the S/N per resolution element of the final spectra; the values listed in Table~\ref{tab:spectroscopy} were computed using a 100-\AA-wide region of the continuum centered at 4600\,\AA.

We fit the six Balmer lines H$\beta$-H9 of our final 1D spectra to pure-hydrogen, 1D model atmospheres for white dwarfs. In short, the models employ the latest treatment of Stark broadening, as well as the ML2/$\alpha=0.8$ prescription of the mixing-length theory. A full description of the models and fitting procedures is contained in \citet{2011ApJ...730..128T}. For each individual DAV, the models were convolved to match the mean seeing during the observations.

The 1D effective temperatures and surface gravities found from these fits are detailed in Table~\ref{tab:spectroscopy}; the fits are visualized in Figure~\ref{fig:balmerfits}. We add in quadrature systematic uncertainties of 1.2\% on the effective temperature and 0.038\,dex on the surface gravity \citep{2005ApJS..156...47L}. We use the analytic functions based on three-dimensional (3D) convection simulations of \citet{2013A&A...559A.104T} to correct these 1D values to 3D atmospheric parameters, which we list in Table~\ref{tab:spectroscopy} as ${T}_{\mathrm{eff}}$-3D and $\log{g}$-3D. We adopt these 3D-corrected atmospheric parameters as our estimates of the effective temperature and surface gravity of our 27 DAVs here, and use them in concert with the white dwarf models of \citet{2001PASP..113..409F} (which have thick hydrogen layers and uniformly mixed, 50\% carbon and 50\% oxygen cores) to estimate the overall mass of our DAVs. Our overall mass estimates do not change by more than the adopted uncertainties (0.04\,\msun\ for all white dwarfs here) if we instead use the models of \citet{2010ApJ...717..183R}, which have more realistic carbon-oxygen profiles resulting from full evolutionary sequences.

Notably, two DAVs in our sample have line-of-sight dM companions, one of which strongly contaminates the H$\beta$ line profiles. For this DAV (EPIC\,220204626, SDSSJ0111+0009) we have omitted the H$\beta$ line from our fits.

Our fitting methodology differs slightly from that adopted by \citet{2016MNRAS.457.2855G} in that we fit the averaged spectrum rather than each spectrum individually, so our reported parameters are slightly different than reported there, but still within the 1$\sigma$ uncertainties. However, we note that the values for KIC\,4357037 (KISJ1917+3927) were misreported in \citet{2016MNRAS.457.2855G}, and are corrected here.

We put our spectroscopic effective temperatures and surface gravities in the context of the empirical DAV instability strip in Figure~\ref{fig:istrip}; we show the empirical blue and red edges found by \citet{2015ApJ...809..148T}. All atmospheric parameters, including literature DAVs and white dwarfs not observed to vary, have been determined using the same 1D models using ML2/$\alpha$=0.8, and all have been corrected to account for the 3D-dependence of convection. Since {\em Kepler} is often sensitive to lower-amplitude pulsations than from the ground, it appears the empirical blue edge of the instability strip may need adjusting by roughly 200\,K to hotter temperatures, but we leave that discussion to future analyses.

We adopt in Figure~\ref{fig:istrip} a color code that we will use throughout the manuscript. We mark in blue all DAVs with a weighted mean period (WMP) shorter than 600\,s (as described in more detail in Section~\ref{sec:wmp}), which cluster at the hotter (blue) half of the instability strip. We mark in gold all DAVs with a WMP exceeding 600\,s. Finally, we mark in red the first six outbursting DAVs \citep{2017ASPC..509..303B}, all of which we mark in Table~\ref{tab:spectroscopy} with a $^{\dagger}$ symbol. These outbursting DAVs are also all cooler than $11{,}300$\,K and exhibit stochastic, large-amplitude flux excursions we have called outbursts.

Outbursts in the coolest DAVs were first discovered in the long-baseline monitoring of the {\em Kepler} spacecraft \citep{2015ApJ...809...14B} and confirmed in a second DAV observed in {\em K2} Campaign 1 \citep{2015ApJ...810L...5H}. Detailed discussion of this interesting new phenomenon is outside the scope of this paper, but we have previously demonstrated that outbursts affect (and may be affected by) pulsations \citep{2015ApJ...810L...5H}.

Interestingly, there are three {\em K2} pulsating white dwarfs near or even cooler than the first six outbursting DAVs. One of these three is GD~1212, which only has 9 days of observations, so the limits on a lack of outbursts are not robust. GD~1212 was re-observed by {\em K2} for more than 75 days in Campaign 12, and will be analyzed in a future publication. We will discuss our insights from {\em K2} into the evolution of white dwarfs through the DAV instability strip in Section~\ref{sec:wmp}.

\section{A Dichotomy of Mode Linewidths}
\label{sec:hwhm}

\begin{figure*}
\centering{\includegraphics[width=0.995\textwidth]{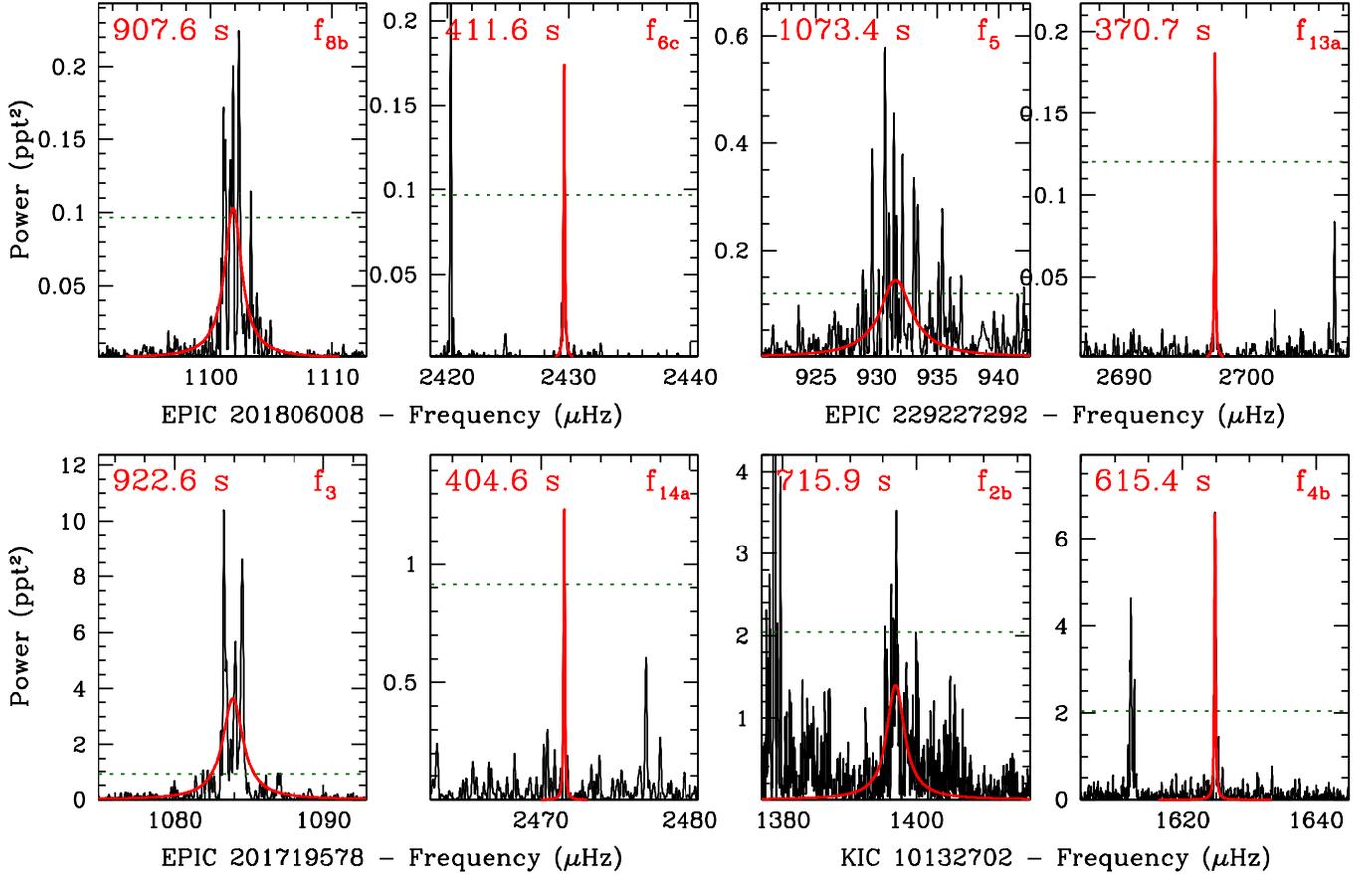}}
\caption{We show detailed portions of the Fourier transforms of four DAVs observed with {\em K2} to show that pulsations with periods exceeding roughly 800\,s (frequencies $<$1250\,\muhz) have power broadly distributed outside the spectral window of the {\em Kepler} observations. For every DAV we have fit a Lorentzian to each independent mode, as described in Section~\ref{sec:hwhm}. At top left we show two modes in EPIC\,201806008 (PG\,1149+058): $f_{8b}$, likely the $m=0$ component of an $\ell=2$ mode centered at 907.6\,s, has power broadly distributed, with HWHM=0.868\,\muhz, whereas $f_{6c}$, the $m=+1$ component of the $\ell=1$ mode centered at 412.40\,s, has a sharp peak with HWHM=0.056\,\muhz. Similarly, in the top right panels we show two modes in EPIC\,229227292 (ATLASJ1342$-$0735): the band of power centered at 1073.44\,s, which has HWHM=1.632\,\muhz, is significantly broader than the $m=-1$ component of the $\ell=1$ mode centered at 370.05\,s. The bottom left panels show similar behavior in EPIC\,201719578 (WD\,1119+042): the 922.60\,s mode in the same star has a HWHM of 0.836\,\muhz, whereas the 404.60\,s mode has a HWHM of 0.045\,\muhz. Finally, we show KIC\,10132702 (KISJ1913+4709) in the bottom right panel, where the 715.86\,s mode has a HWHM of 1.632\,\muhz, whereas the 615.44\,s mode has a HWHM of 0.104\,\muhz. The x-axis (frequency) scales are identical for each star. The best fits for each mode are detailed in Table~\ref{tab:pulsations} and over-plotted here; the HWHM of Lorentzian fits to all 27 DAVs are shown in Figure~\ref{fig:periodhwhm}. \label{fig:lorcomp}}
\end{figure*}

Long-baseline {\em Kepler} photometry has opened an unprecedented window into the frequency and amplitude stability of DAVs. From early observations of the first DAV observed with {\em Kepler}, we noticed broad bands of power in the Fourier transform. Rather than just a small number of peaks, we saw a large number of peaks under a broad envelope; the envelope was significantly wider than the sharp spectral windows \citep{2015ApJ...809...14B}. We saw similarly broad power bands in another cool DAV, GD\,1212, observed during an engineering test run for {\em K2} \citep{2014ApJ...789...85H}. The broadly distributed power was reminiscent of the power spectra of stochastically driven oscillations in the Sun or other solar-like oscillators \citep{2013ARA&A..51..353C}. However, as noted in \citet{2015ApJ...809...14B}, white dwarfs have a sound-crossing time orders of magnitude shorter than their observed pulsation periods, so these broad linewidths cannot be related to stochastic driving.

Additionally, it became clear that not all white dwarfs with long-baseline, space-based observations showed such broad linewidths. For example, we observed multiple DAVs with hotter effective temperatures and shorter-period pulsations that were completely coherent, within the uncertainties, over several months of {\em Kepler} observations \citep{2014MNRAS.438.3086G,2015MNRAS.451.1701H}.

Within the larger sample of the first 27 DAVs, we have noticed an emergent pattern: a dichotomy of mode linewidths, even within the same star, delineated almost entirely by the pulsation period. The shorter-period modes appear stable in phase, producing narrow peaks in the Fourier transform, while the longer-period modes are phase unstable, spreading their power over a broad band in the Fourier transform. To quantify this behavior we have fit Lorentzian envelopes to every significant pulsation period, and compare the period determined to the half-width-at-half-maximum (HWHM, $\gamma$) of the Lorentzian fits. Motivated by \citet{1994A&A...289..649T}, we have fit each thicket of peaks in the power spectrum (the square of the Fourier transform) with the function
\begin{equation}
{\mathrm{Power}} = \frac{A \gamma^2}{(\nu - \nu_0)^2 + \gamma^2} + B
\end{equation}
where $A$ represents the Lorentzian height, $\gamma$ is the HWHM, $\nu_0$ is the central frequency of the peak, and $B$ represents a DC offset for the background, which we hold fixed for all modes in the same star as the median power of the entire power spectrum. We have only fit the region within 5\,\muhz\ to each side of the highest-amplitude peak within each band of power; this highest peak defines our initial guess for the central frequency, and we use an initial guess for the HWHM of 0.2\,\muhz, roughly twice our typical frequency resolution.

\begin{figure*}
\centering{\includegraphics[width=0.85\textwidth]{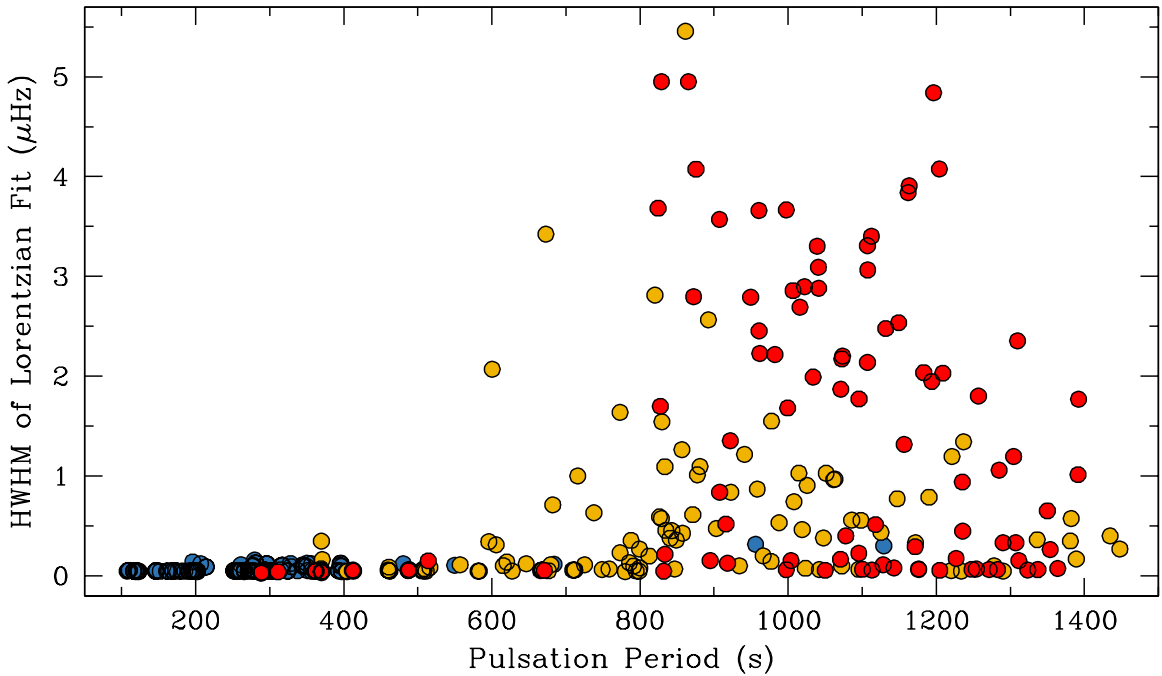}}
\caption{The half width at half maximum (HWHM) of Lorentzian functions fit to all significant peaks in the power spectra of the 27 DAVs observed through {\em K2} Campaign 8 by the {\em Kepler} space telescope; our procedure is described in Section~\ref{sec:hwhm}. We use the same color classification as in Figure~\ref{fig:istrip}, where blue are objects with WMP $<$600\,s, gold with WMP $>$600\,s, and red those with outbursts (see Section~\ref{sec:wmp}). We have excluded any nonlinear combination frequencies in this analysis. We see a sharp increase in HWHM at roughly 800\,s, indicating that modes with relatively high radial order ($k>15$ for $\ell=1$ modes) are not coherent in phase, similar in behavior to stochastically excited pulsators. We save discussion of the possible physical mechanisms behind this phenomenon for future work (Montgomery et al., in prep.). \label{fig:periodhwhm}}
\end{figure*}

We illustrate the nature of the dichotomy of mode linewidths in Figure~\ref{fig:lorcomp}. Here we show, for four different DAVs in our sample, two independent modes within the same star that are well separated in period. In each case, the shorter-period modes have linewidths roughly matching the spectral window of the observations, with HWHM $\leq$0.1\,\muhz. All DAVs in Figure~\ref{fig:lorcomp} were observed by the {\em Kepler} spacecraft for more than 78\,days with a duty cycle exceeding 91\%. We observe that modes exceeding roughly 800\,s (with frequencies below roughly 1250\,\muhz) have much broader HWHM, often exceeding 1\,\muhz.

These broad bands of power are most likely representative of a single mode that is unstable in phase, reminiscent of a damped harmonic oscillator. For example, in the top left panels of Figure~\ref{fig:lorcomp} we show two modes in EPIC\,201806008 (PG\,1149+058), the second outbursting DAV discovered \citep{2015ApJ...810L...5H}. The mode $f_{6c}$ is the $m=+1$ component of the $\ell=1$ mode centered at 412.40\,s --- the two other components are seen at slightly lower frequency (we discuss this splitting in the context of rotation in Section~\ref{sec:rotation}). The HWHM of $f_{6c}$ is consistent with the spectral window: $\gamma=0.056$\,\muhz. At much lower frequency, we identify the mode $f_{8b}$ as the likely $m=0$ component of an $\ell=2$ mode centered at 907.58\,s. It has power much more broadly distributed, with $\gamma=0.868$\,\muhz, too narrow to encompass other normal modes of different degree or radial order, but much broader than the spectral window.

We extend this Lorentzian analysis of mode linewidths to the entire sample of 27 DAVs in Figure~\ref{fig:periodhwhm}, which shows the period and HWHM computed for every Lorentzian fit to all 225 significant pulsations (we exclude all nonlinear combination frequencies from Figure~\ref{fig:periodhwhm}; see Section~\ref{sec:wmp}). We performed the same experiment by fitting Gaussians to each peak and obtained a nearly identical distribution, so our choice of function does not significantly alter our results.

We see a sharp increase in HWHM at roughly 800\,s, indicating that modes with relatively high radial order ($k>15$ for $\ell=1$ modes) are not phase coherent, exhibiting behavior similar to stochastically excited pulsators, though we do not believe the mode instability is related to stochastic driving.

One possibility is that the broad linewidths arise from a process that has disrupted the phase coherence of the modes. In analogy with stochastically-driven modes, we relate the observed mode linewidth (HWHM) to its lifetime ($\tau$) via the relation $\tau=2/(\pi \gamma)$ \citep{2009A&A...500L..21C}. Doing so for the longer-period modes ($\gamma>1$\,\muhz), we find mode lifetimes of order several days to weeks. This is comparable to the damping rates we expect for roughly 1000\,s $\ell=1$ modes of typical pulsating white dwarfs \citep{1999ApJ...511..904G}. We will explore possible physical mechanisms to explain these broad linewidths in a future publication (Montgomery et al., in prep.).

Determining a pulsation period to use for asteroseismology of the coolest DAVs with the longest-period modes is different than for the hot DAVs with narrow pulsation peaks that are determined via a linear least-squares fit and cleanly prewhiten. Therefore, for all modes with fitted HWHM exceeding 0.2\,\muhz, we do not include their linear-least-square fits in Table~\ref{tab:pulsations}, and instead only include the periods and amplitudes as determined from the Lorentzian fits in Table~\ref{tab:pulsations}, with period uncertainties computed from the HWHM.

\section{Characteristics of the DAV Instability Strip}
\label{sec:wmp}

\begin{figure*}
\centering{\includegraphics[width=0.85\textwidth]{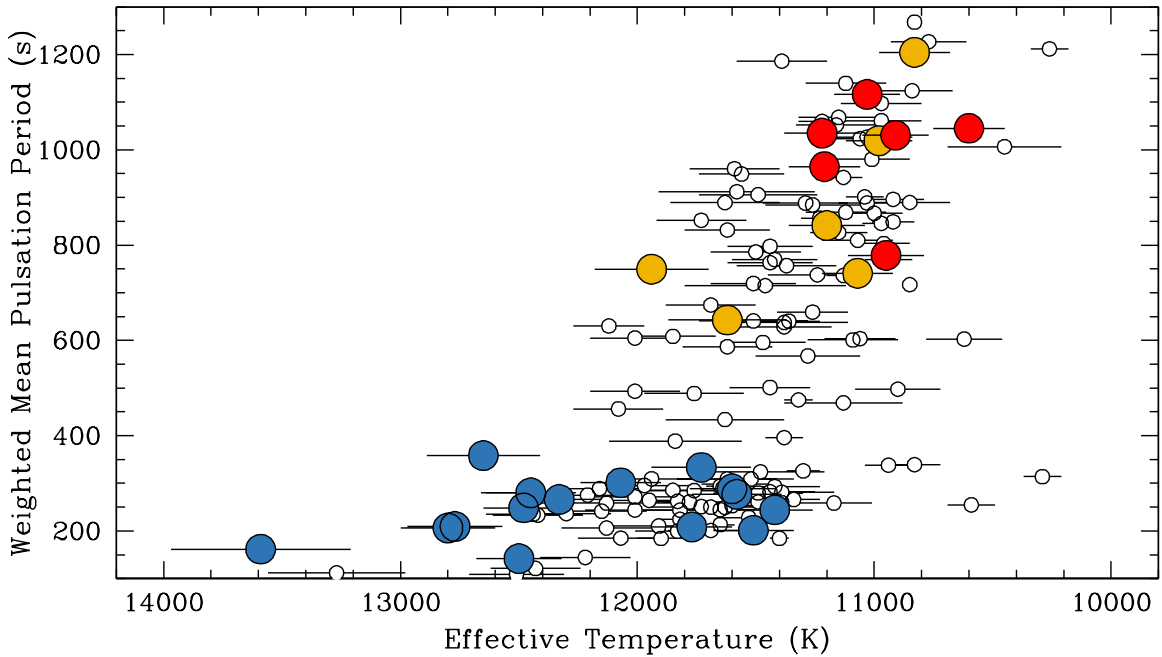}}
\caption{Following \citet{1993BaltA...2..407C}, we plot the effective temperature versus the weighted mean pulsation period (WMP) of each DAV analyzed here, using the same color classification used in Figure~\ref{fig:istrip}. We also show in white circles the 114 known DAVs from the literature with spectroscopy analyzed in the same way, also plotted in Figure~\ref{fig:istrip}. \label{fig:wmpteff}}
\end{figure*}

\begin{figure*}[t]
\centering{\includegraphics[width=0.995\textwidth]{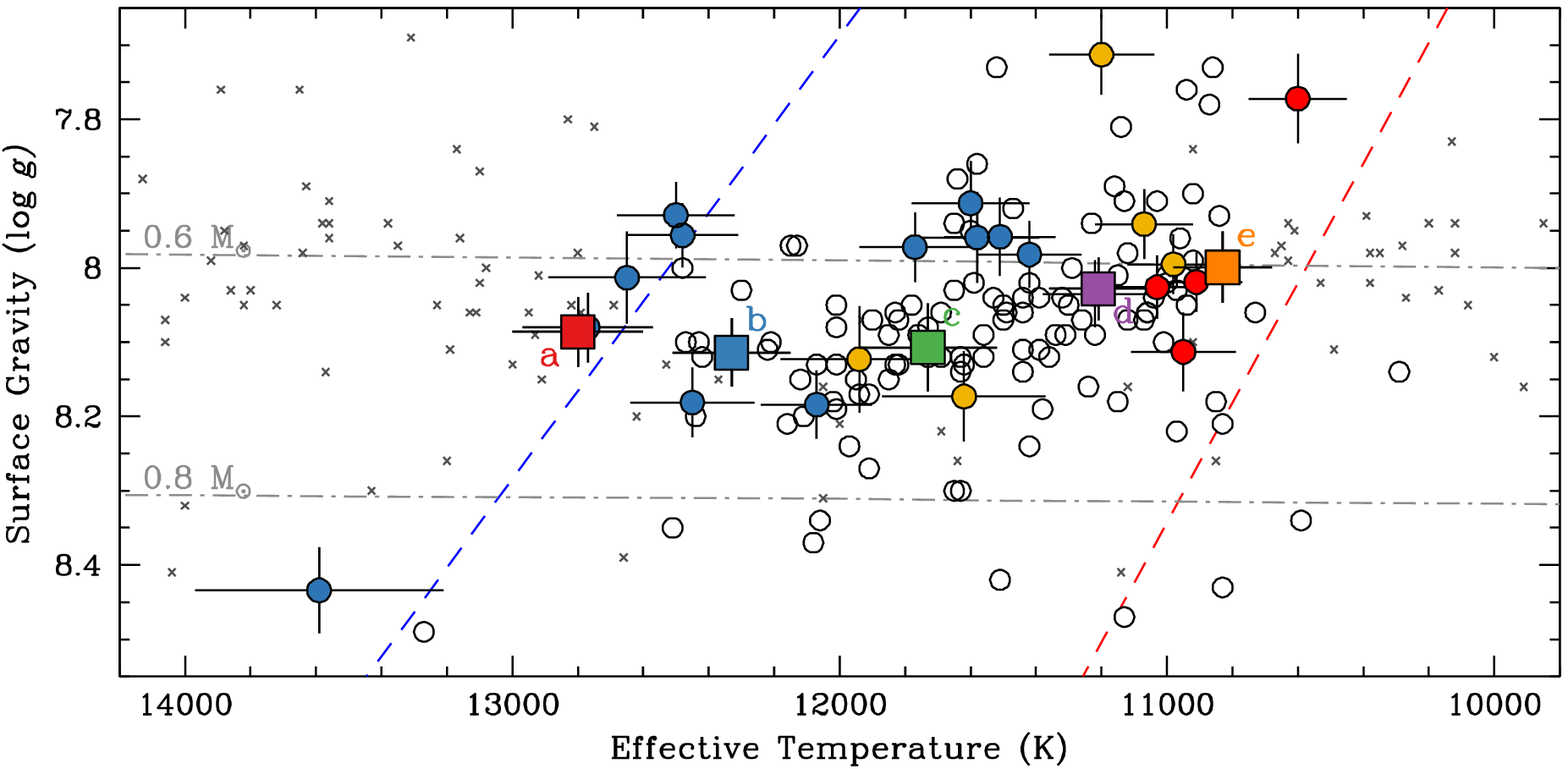}
\includegraphics[width=0.995\textwidth]{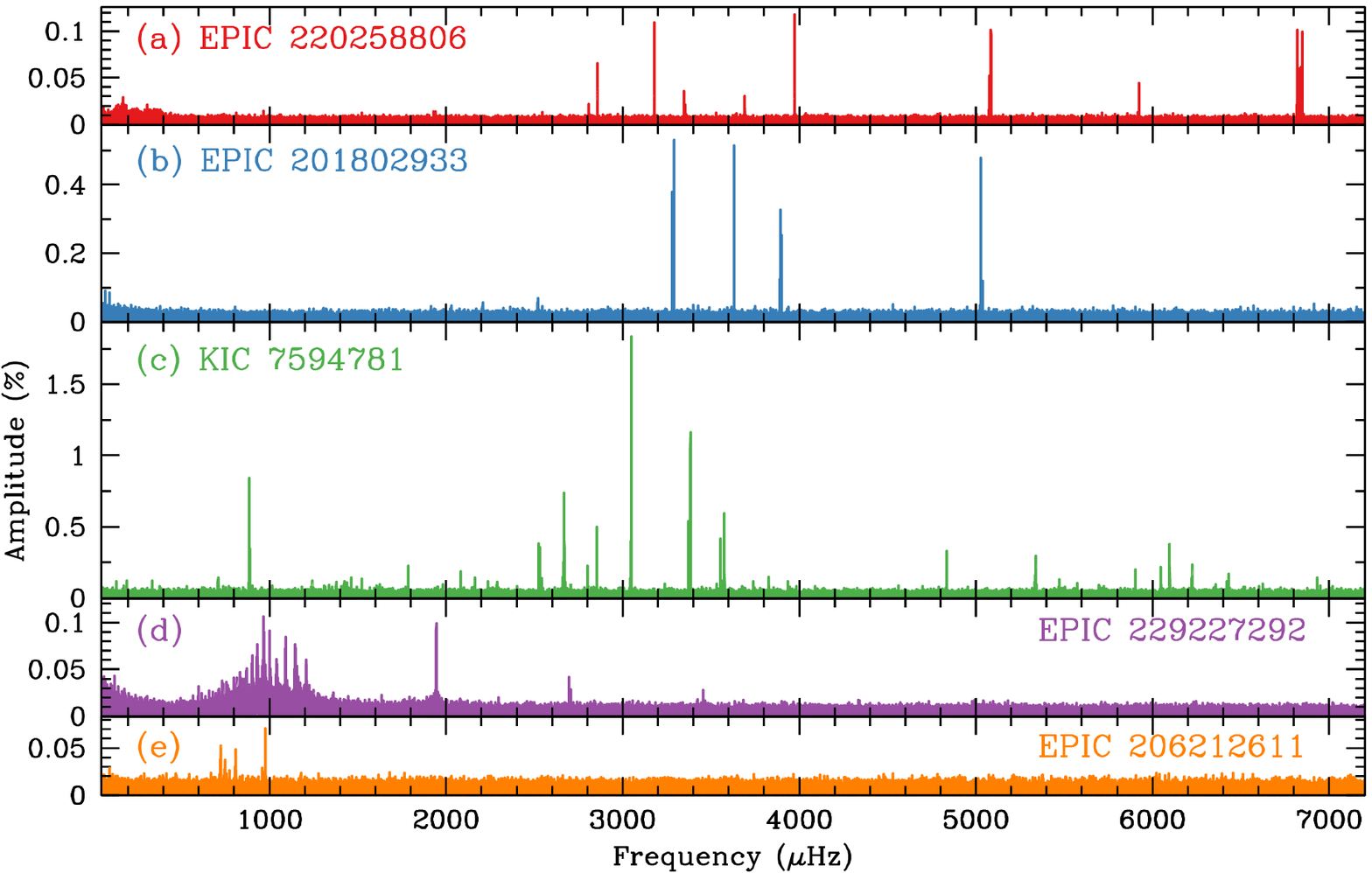}}
\caption{The top panel shows the same information as Figure~\ref{fig:istrip} but color codes five representative stars in our sample to highlight the various phases of DAV cooling illuminated by {\em Kepler} observations so far, especially pulsation period and amplitude evolution. In order of decreasing effective temperature, we highlight: (a) EPIC 220258806 (SDSSJ0106+0145), (b) EPIC 201802933 (SDSSJ1151+0525), (c) KIC 7594781 (KISJ1908+4316), (d) EPIC 229227292 (ATLASJ1342$-$0735), and (e) EPIC 206212611 (SDSSJ2220$-$0933). Fourier transforms of each are shown below with the same order and color coding, with amplitudes adjusted to reflect target flux fractions found in Table~\ref{tab:kepler}.
DAVs begin pulsating at the blue edge of the instability strip (a) with low-$k$ (1<$k$<4) $\ell=1,2$ modes from roughly $100-300$\,s and relatively low amplitudes ($\sim$1\,ppt). As they cool (b), their convection zones deepen, driving longer-period (lower-frequency) pulsations with growing amplitudes. DAVs in the middle of the instability strip (c) have high-amplitude modes and the greatest number of nonlinear combination frequencies. As they cool further (d), it appears common for DAVs to undergo aperiodic, sporadic outbursts; these DAVs tend to have many low-frequency modes excited in addition to one or several shorter-period, stable pulsations. Finally, a handful of DAVs have effective temperatures cooler than the outbursting DAVs but do not experience large-scale flux excursions (e); these coolest DAVs have the longest-period pulsations with relatively low amplitudes. \label{fig:colorfts}}
\end{figure*}

Pulsations in DAVs are excited by the ``convective driving'' mechanism, which is intimately tied to the surface hydrogen partial-ionization zone that develops as a white dwarf cools below roughly $13{,}000$\,K \citep{1991MNRAS.251..673B,1999ApJ...519..783W}. Driving is strongest for pulsations with periods nearest to roughly 25 times the thermal timescale at the base of the convection zone \citep{1999ApJ...511..904G}.

By the time white dwarfs reach the DAV instability strip, their evolution is dominated by secular cooling. As they cool, the surface convection zone deepens, driving longer-period pulsations (e.g., \citealt{2013ApJ...762...57V}). These ensemble characteristics of the DAV instability strip, especially the lengthening of periods with cooler effective temperature, have been considered for decades (e.g., \citealt{1994PhDT........29C}), and were most recently summarized observationally by \citet{2006ApJ...640..956M}.

The most common way to visualize the increasing mode periods as DAVs cool is to compute the weighted mean period (WMP) of the significant pulsations \citep{1993BaltA...2..407C}. The WMP is linearly weighted by the amplitude of each mode, such that WMP = $\sum_{i} P_i A_i / \sum_{i} A_i$, where $P_i$ and $A_i$ are the period and amplitude, respectively, of each significant independent pulsation detected (excluding nonlinear combination frequencies). We show the WMPs of all 27 DAVs observed by {\em Kepler} and {\em K2} so far in Figure~\ref{fig:wmpteff}, in addition to the WMP and temperatures of previously known DAVs with atmospheric parameters determined in the same way, by \citet{2011ApJ...743..138G} and \citet{2011ApJ...730..128T}.

Extensive coverage by the {\em Kepler} space telescope affords an even more nuanced exploration of the DAV instability strip. With our follow-up spectroscopic characterization of the first 27 DAVs observed with {\em Kepler} and {\em K2}, we can effectively put the evolutionary state of our pulsating white dwarfs into greater context. Figure~\ref{fig:colorfts} highlights five DAVs that are representative of five different stages of evolution through the DAV instability strip, including color-coded Fourier transforms of each representative object. These data reveal important trends in the period and amplitude evolution of DAVs through the instability strip. Since there are likely mass-dependent effects, we describe this sequence for DAVs near the mean field white dwarf mass of 0.62\,\msun, in order of hottest to coolest DAVs:

(a) DAVs begin pulsating at the blue edge of the instability strip with low-$k$ ($1<k<6$) $\ell=1,2$ modes from roughly $100-300$\,s and relatively low amplitudes ($\sim$1\,ppt). We highlight in red in Figure~\ref{fig:colorfts} the hot ($12{,}800$\,K) white dwarf EPIC 220258806 (SDSSJ0106+0145) that appears to fall at the blue edge of the instability strip. This DAV has stable pulsations ranging from $116.28-356.14$\,s (WMP = 206.0\,s), most of which are clearly identified as both $\ell=1$ and $\ell=2$ modes (see Table~\ref{tab:pulsations}), although no modes have amplitudes in excess of roughly 1\,ppt. The hottest DAVs in our sample have relatively low-amplitude modes. For example, the hottest DAV currently known is EPIC 211914185 (SDSSJ0837+1856) which has just two relatively low-amplitude ($1-3$\,ppt) dipole modes centered at 109.15\,s and 190.45\,s \citep{2017ApJ...841L...2H}. An example of a blue-edge DAV in this phase of evolution well-studied from ground-based observations is G226$-$29 \citep{1995ApJ...447..874K}.

(b) As DAVs cool by a few hundred degrees from the blue edge, they retain relatively short-period pulsations but their observed amplitudes increase. For example, the $12{,}330$\,K EPIC 201802933 (SDSSJ1151+0525) has amplitudes up to roughly 5\,ppt with modes ranging from $123.11-396.62$\,s (WMP = 266.2\,s). These low-$k$ modes are expected to have extremely long mode lifetimes \citep{1999ApJ...511..904G}, and most modes in these hot DAVs with mode periods shorter than 400\,s appear coherent in phase. A well-known example of this phase of DAV evolution is G117-B15A \citep{1995ApJ...438..908R,2000ApJ...534L.185K}.

\begin{figure*}
\centering{\includegraphics[width=0.995\textwidth]{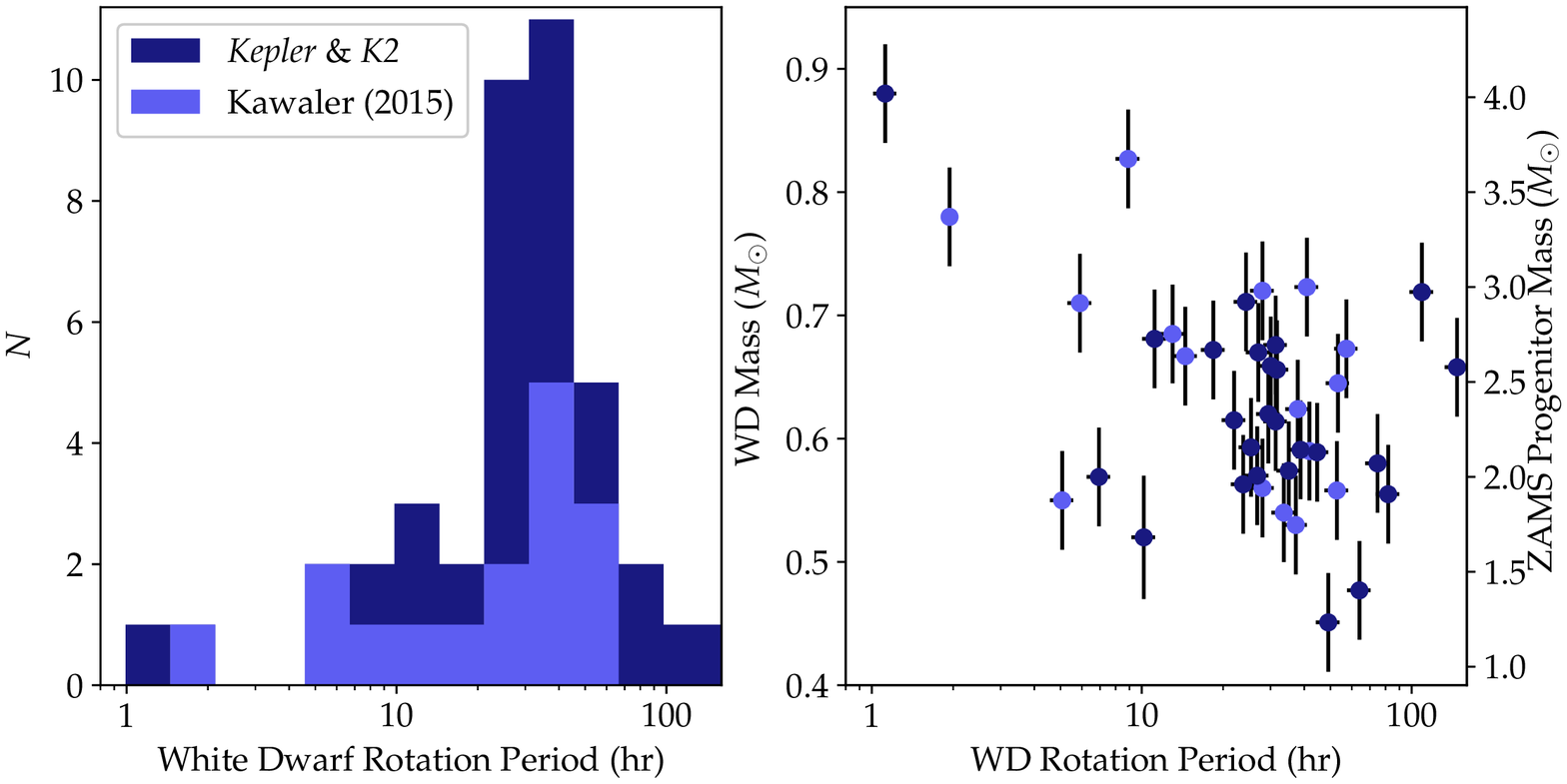}}
\caption{We compare asteroseismically determined rotation periods for all known pulsating white dwarfs, detailed in Table~\ref{tab:rotation}. All white dwarfs presented here appear to be isolated stars, so these rotation periods should be representative of the endpoints of single-star evolution; we have excluded the only known close binary (a WD+dM in a 6.9-hr orbit), EPIC 201730811 (SDSSJ1136$+$0409, \citealt{2015MNRAS.451.1701H}). The left histogram shares the color-coding of the right panel, which compares white dwarf rotation as a function of mass. Estimates of the ZAMS progenitor masses for each white dwarf are listed on the right axis. Notably, EPIC 211914185 (SDSSJ0837+1856) is more massive ($0.88\pm0.03$\,\msun) and rotates faster ($1.13\pm0.02$\,hr) than any other pulsating white dwarf \citep{2017ApJ...841L...2H}; we see evidence for a link between high mass and fast rotation, but require additional massive white dwarfs to confirm this trend. \label{fig:rotation}}
\end{figure*}

(c) After DAVs cool by nearly 1000\,K from the blue edge of the instability strip, their dominant pulsations exceed 300\,s. DAVs in the middle of the instability strip tend to have very high-amplitude modes and the greatest number of nonlinear combination frequencies. For example, KIC 7594781 (KISJ1908+4316) is an $11{,}730$\,K white dwarf with the largest number of pulsation frequencies presented in our sample (WMP = 333.3\,s). However, fewer than 10 of these frequencies of variability actually arise from unstable stellar oscillations: the rest are nonlinear combination frequencies (e.g., peaks corresponding to $f_1$+$f_2$ or 2$f_1$), which arise from distortions of a linear pulsation signal caused by the changing convection zone of a DAV \citep{1992MNRAS.259..519B,2005ApJ...633.1142M}. These are not independent modes but rather artifacts of some nonlinear distortion in the outer layers of the star \citep{2013ApJ...766...42H}.

Objects in the middle of the DAV instability strip represent a transition from hot, stable pulsations to cool DAVs with oscillations that are observed to change in frequency and/or amplitude on relatively short timescales. G29$-$38 is a prominent, long-studied example of DAV representative of this phase in the middle of the instability strip \citep{1998ApJ...495..424K}.

(d) {\em Kepler} observations have now confirmed what appears to be a new phase in the evolution of DAVs as they approach the cool edge of the instability strip: aperiodic outbursts that increase the mean stellar flux by a few to 15\%, last several hours, and recur sporadically on a timescale of days. Outbursts were first discovered in the original {\em Kepler} mission \citep{2015ApJ...809...14B}, and quickly confirmed in {\em K2} \citep{2015ApJ...810L...5H}. Observational properties of the first six outbursting DAVs are detailed in \citet{2017ASPC..509..303B}. Aside from many long-period pulsations, almost all of the outbursting DAVs observed by {\em Kepler} also show at least one and sometimes multiple shorter-period modes. Figure~\ref{fig:colorfts} highlights the Fourier transform of the $11{,}210$\,K outbursting DAV EPIC 229227292 (ATLASJ1342$-$0735), which is dominated by pulsations from $827.7-1323.6$\,s (WMP = 964.5\,s) but also shows significant periods centered around 514.1\,s, 370.0\,s and 288.9\,s (see Table~\ref{tab:pulsations}). From the brightest outbursting DAV observed so far (PG\,1149+058) we saw strong evidence that the pulsations respond to the outbursts \citep{2015ApJ...810L...5H}.

(e) Finally, a handful of DAVs have effective temperatures as cool or cooler than the outbursting DAVs but do not experience large-scale flux excursions, suggesting that not all DAVs outburst at the cool edge of the instability strip. The coolest DAVs tend to have the longest-period pulsations with relatively low amplitudes. We highlight in Figure~\ref{fig:colorfts} the star EPIC 206212611 (SDSSJ2220$-$0933), which is one of the coolest DAVs known, at $10{,}830$\,K. We do not detect any outbursts to a limit of 5.5\%, following the methodology outlined in \citet{2016ApJ...829...82B}. This DAV has modes from $1023.4-1381.2$\,s (WMP = 1204.4\,s), all at relatively low amplitudes below $1$\,ppt.

\section{Rotation Rates}
\label{sec:rotation}

\begin{deluxetable*}{llccccccccc}
\tabletypesize{\scriptsize}
\tablecolumns{11}
\tablewidth{0pc}
\tablecaption{Rotation rates of isolated white dwarfs determined via asteroseismology. \label{tab:rotation}}
\tablehead{
        \colhead{Star Name} &
        \colhead{RA \& Dec (J2000)} &
        \colhead{Class} &
        \colhead{\teff-3D} &
        \colhead{WD Mass} &
        \colhead{$\Delta M_{\mathrm{WD}}$} &  
        \colhead{Prog. Mass} &
        \colhead{$\Delta M_{\mathrm{Prog.}}$} & 
        \colhead{$\delta \nu_{\ell=1}$} &  
        \colhead{$P_{\mathrm{rot}}$} &  
        \colhead{Ref.} \\
  &  &  & (K) & (\msun) & (\msun) & (\msun) & (\msun) & (\muhz) & (hr) & }
\startdata
\multicolumn{11}{c}{Literature values for all classes of pulsating white dwarfs compiled by \citet{2015ASPC..493...65K}} \\
\hline   
 Ross 548  & 01 36 13.58 $-$11 20 32.7  & DAV & 12300 & 0.62 & 0.04 & 2.36 & 0.47 & \nodata & 37.8 & 1 \\
 HL Tau 76  & 04 18 56.63 +27 17 48.5  & DAV & 11470 & 0.56 & 0.04 & 1.93 & 0.44 & \nodata & 52.8 & 2 \\
 HS 0507+0434 & 05 10 13.94 +04 38 38.4  & DAV & 12010 & 0.72 & 0.04 & 3.00 & 0.52 & 3.6 & 40.9 & 3 \\
 KUV11370+4222  & 11 39 41.42 +42 05 18.7  & DAV & 11940 & 0.71 & 0.04 & 2.92 & 0.51 & 25.0 & 5.9 & 4 \\
 GD 154  & 13 09 57.69 +35 09 47.1  & DAV & 11120 & 0.65 & 0.04 & 2.49 & 0.48 & 2.8 & 53.3 & 5 \\
 LP 133$-$144 & 13 51 20.24 +54 57 42.6  & DAV & 12150 & 0.59 & 0.04 & 2.13 & 0.45 & \nodata & 41.8 & 6 \\
 GD 165  & 14 24 39.14 +09 17 14.0  & DAV & 12220 & 0.67 & 0.04 & 2.68 & 0.50 & \nodata & 57.3 & 1 \\
 L19-2  & 14 33 07.60 $-$81 20 14.5  & DAV & 12070 & 0.69 & 0.04 & 2.75 & 0.50 & 13.0 & 13.0 & 7 \\
 SDSSJ1612+0830 & 16 12 18.08 +08 30 28.1  & DAV & 11810 & 0.78 & 0.04 & 3.50 & 0.55 & 75.6 & 1.9 & 8 \\
 G226-29  & 16 48 25.64 +59 03 22.7  & DAV & 12510 & 0.83 & 0.04 & 3.68 & 0.57 & \nodata & 8.9 & 9 \\
 G185-32  & 19 37 13.68 +27 43 18.7  & DAV & 12470 & 0.67 & 0.04 & 2.64 & 0.49 & \nodata & 14.5 & 10 \\
 PG 0122+200 & 01 25 22.52 +20 17 56.8  & DOV & 80000 & 0.53 & 0.04 & 1.75 & 0.42 & 3.7 & 37.2 & 11 \\
 NGC 1501 & 04 06 59.39 +60 55 14.4  & DOV & 134000 & 0.56 & 0.04 & 1.94 & 0.44 & \nodata & 28.0 & 12 \\
 PG 1159$-$035 & 12 01 45.97 $-$03 45 40.6  & DOV & 140000 & 0.54 & 0.04 & 1.81 & 0.43 & \nodata & 33.6 & 13 \\
 RX J2117.1+3412 & 21 17 08.28 +34 12 27.5  & DOV & 170000 & 0.72 & 0.04 & 2.98 & 0.52 & \nodata & 28.0 & 14 \\
 PG 2131+066 & 21 34 08.23 +06 50 57.4  & DOV & 95000 & 0.55 & 0.04 & 1.88 & 0.43 & \nodata & 5.1 & 15 \\
KIC 8626021 & 19 29 04.69 +44 47 09.7  & DBV & 30000 & 0.59 & 0.04 & 2.10 & 0.45 & 3.3 & 44.6 & 16 \\
EPIC  220670436 & 01 14 37.66 +10 41 04.8  & DBV & 31300 & 0.52 & 0.04 & 1.68 & 0.42 & 15.4 & 10.2 & 17 \\
\hline
\multicolumn{11}{c}{Rotation rates from DAVs analyzed here} \\
\hline                     
KIC  4357037 & 19 17 19.20 +39 27 19.1 & DAV & 12650 & 0.62 & 0.04 & 2.30 & 0.47 & 6.7 & 22.0 & 0 \\
KIC  4552982 & 19 16 43.83 +39 38 49.7 & DAV & 10950 & 0.67 & 0.04 & 2.67 & 0.50 & 8.0 & 18.4 & 18 \\
KIC  7594781 & 19 08 35.88 +43 16 42.4 & DAV & 11730 & 0.67 & 0.04 & 2.66 & 0.49 & 5.5 & 26.8 & 0 \\
KIC  10132702 & 19 13 40.89 +47 09 31.3 & DAV & 11940 & 0.68 & 0.04 & 2.73 & 0.50 & 13.2 & 11.2 & 0 \\
KIC  11911480 & 19 20 24.90 +50 17 21.3 & DAV & 11580 & 0.58 & 0.04 & 2.07 & 0.45 & 1.97 & 74.7 & 19 \\
EPIC  60017836 & 23 38 50.74 $-$07 41 19.9 & DAV & 10980 & 0.57 & 0.04 & 2.00 & 0.44 & 21.2 & 6.9 & 0 \\
EPIC  201719578 & 11 22 21.10 +03 58 22.4 & DAV & 11070 & 0.57 & 0.04 & 2.01 & 0.44 & 5.5 & 26.8 & 0 \\
EPIC  201730811 & 11 36 55.16 +04 09 52.8 & DAV & 12480 & 0.58 & 0.04 & 2.08 & 0.45 & 56.7 & 2.6 & 20 \\
EPIC  201802933 & 11 51 26.15 +05 25 12.9 & DAV & 12330 & 0.68 & 0.04 & 2.69 & 0.50 & 4.7 & 31.3 & 0 \\
EPIC  201806008 & 11 51 54.20 +05 28 39.8 & DAV & 10910 & 0.61 & 0.04 & 2.29 & 0.47 & 4.7 & 31.3 & 0 \\
EPIC  210397465 & 03 58 24.23 +13 24 30.8 & DAV & 11200 & 0.45 & 0.04 & 1.23 & 0.38 & 3.0 & 49.1 & 0 \\
EPIC  211596649 & 08 32 03.98 +14 29 42.4 & DAV & 11600 & 0.56 & 0.04 & 1.91 & 0.44 & 1.8 & 81.8 & 0 \\
EPIC  211629697 & 08 40 54.14 +14 57 09.0 & DAV & 10600 & 0.48 & 0.04 & 1.40 & 0.40 & 2.3 & 64.0 & 0 \\
EPIC  211914185 & 08 37 02.16 +18 56 13.4 & DAV & 13590 & 0.88 & 0.04 & 4.02 & 0.60 & 131.6 & 1.1 & 21 \\
EPIC  211926430 & 09 00 41.08 +19 07 14.4 & DAV & 11420 & 0.59 & 0.04 & 2.16 & 0.46 & 5.8 & 25.4 & 0 \\
EPIC  228682478 & 08 40 27.84 +13 20 10.0 & DAV & 12070 & 0.72 & 0.04 & 2.97 & 0.52 & 1.4 & 109.1 & 0 \\
EPIC  229227292 & 13 42 11.62 $-$07 35 40.1 & DAV & 11210 & 0.62 & 0.04 & 2.33 & 0.47 & 5.0 & 29.4 & 0 \\
EPIC  220204626 & 01 11 23.89 +00 09 35.2 & DAV & 11620 & 0.71 & 0.04 & 2.92 & 0.51 & 6.1 & 24.3 & 0 \\
EPIC  220258806 & 01 06 37.03 +01 45 03.0 & DAV & 12800 & 0.66 & 0.04 & 2.58 & 0.49 & 4.9 & 30.0 & 0 \\
EPIC  220347759 & 00 51 24.25 +03 39 03.8 & DAV & 12770 & 0.66 & 0.04 & 2.56 & 0.49 & 4.7 & 31.7 & 0
 \enddata
\tablerefs{ (0) This work; (1) \citet{2016ApJS..223...10G}; (2) \citet{2006A&A...446..237D}; (3) \citet{2013MNRAS.429.1585F}; (4) \citet{2014MNRAS.437.2566S}; (5) \citet{1996A&A...314..182P}; (6) \citet{2016MNRAS.461.4059B}; (7) \citet{2001ApJ...552..326B}; (8) \citet{2013MNRAS.430...50C}; (9) \citet{1995ApJ...447..874K}; (10) \citet{2006A&A...453..219P}; (11) \citet{2007A&A...467..237F}; (12) \citet{1996AJ....112.2699B}; (13) \citet{2009Natur.461..501C}; (14) \citet{2002A&A...381..122V}; (15) \citet{1995ApJ...450..350K}; (16) \citet{2011ApJ...736L..39O}; (17) \citet{2017ApJ...835..277H}; (18) \citet{2015ApJ...809...14B}; (19) \citet{2014MNRAS.438.3086G}; (20) \citet{2015MNRAS.451.1701H} --- excluded from Figure~\ref{fig:rotation} since white dwarf in close, post-common-envelope binary; (21) \citet{2017ApJ...841L...2H}.  }
\end{deluxetable*}

The oscillations observed in white dwarfs are non-radial $g$-modes, which can be decomposed into spherical harmonics. Pulsations can thus be represented by three quantum numbers: the radial order ($k$, often expressed as $n$ in other fields of asteroseismology), representing the number of radial nodes; the spherical degree ($\ell$), representing the number of nodal lines expressed at the surface of the star; and the azimuthal order ($m$), representing the number of nodal lines in longitude at the surface of the star, ranging from $-\ell$ to $\ell$.

Due to strong geometric cancellation of higher-$\ell$ modes, we typically observe $\ell=1,2$ modes in white dwarfs. Rotation causes a lifting of degeneracy in the pulsation frequencies, causing a mode to separate into $2\ell+1$ components in $m$ \citep{1989nos..book.....U}. If all multiplets are present, this can result in triplets (for $\ell=1$ modes) and quintuplets (for $\ell=2$ modes) in the presence of slow rotation. This is corroborated by the many DAVs we observed with {\em Kepler} that show triplets and quintuplets of peaks evenly spaced in frequency: an exquisite example is the hot DBV PG\,0112+104 \citep{2017ApJ...835..277H}. Observationally (and in contrast to solar-like oscillations), white dwarf pulsations do not partition mode energy equally into the various $m$ components, such that the amplitudes of different $m$ components for a given $k,\ell$ are not necessarily symmetric.

Still, we can use the clear patterns of frequency spacing in the Fourier transforms of our DAVs to say something about the rotation of these stellar remnants, summarized in Figure~\ref{fig:rotation}. This task is often complex from the ground due to diurnal aliasing, but the extended {\em Kepler} observations significantly simplifies mode identification and affords us the opportunity to probe internal rotation for 20 of the 27 DAVs we present here. For example, Figure~\ref{fig:ftc1} shows the five dipole ($\ell=1$) modes in EPIC 201802933 (SDSSJ1151+0525) identified from their splittings. Evidence for splittings in other DAVs analyzed here can be found in Section~\ref{sec:individual}.

To first order, the observed pulsation frequency splittings ($\delta \nu$) are related to the stellar rotation frequency ($\Omega$) by the relation
\begin{equation}
\delta \nu = m (1 - C_{k,\ell}) \Omega
\end{equation}
where $C_{k,\ell}$ represents the effect of the Coriolis force on the pulsations, as formulated by \citet{1951ApJ...114..373L}. For high-$k$ modes that approach the asymptotic, mode-independent limit for $\ell=1$ modes, $C_{k,1} \to 0.50$. However, the low-$k$ (especially $k<10$) modes found in hot DAVs can suffer from strong mode-trapping effects that also significantly alter $C_{k,\ell}$ (e.g., \citealt{1992ApJS...80..369B}). Theoretical DAV models in, for example, \citet{2012MNRAS.420.1462R}, predict $C_{k,1}$ values typically ranging from roughly $0.45-0.49$, with exceptions down to $C_{k,1}\sim0.35$ for strongly trapped modes.

For the 20 DAVs here for which we have identified modes, all have at least one and many have multiple sets of $\ell=1$ modes; the full catalog of asteroseismic rotation rates for white dwarfs is detailed in Table~\ref{tab:rotation}. To provide the best model-independent estimates for the rotation periods of the newly identified DAVs here, we compute the median frequency splitting of all identified $\ell=1$ modes ($\delta \nu_{\ell=1}$), which we list in Table~\ref{tab:rotation}. We then hold fixed $C_{k,1}=0.47$ and use $\delta \nu_{\ell=1}$ to estimate the DAV rotation period. We put these new DAV rotation periods into the context of all other white dwarfs with measured rotation periods from asteroseismology compiled by \citet{2015ASPC..493...65K} in Figure~\ref{fig:rotation}. {\em Kepler} and {\em K2} have more than doubled the number of white dwarfs for which we have measured internal rotation from pulsations.

We note that it is necessary to perform a complete asteroseismic analysis in order to estimate the best values for $C_{k,\ell}$ for the pulsation modes presented here. This also affords the opportunity to significantly improve constraints on the rotation period; the choice of $C_{k,\ell}$ for each mode dominates the uncertainty on the overall rotation period measured from the frequency splittings, since mode trapping is very common. Given our method of uniformly adopting $C_{k,1}=0.47$, we estimate each rotation period has a systematic uncertainty of roughly 10\%, though we note that $C_{k,\ell}$ cannot exceed 0.50.

Importantly, because we have measured the atmospheric parameters of all DAVs observed with {\em Kepler} so far, we have for the first time a large enough sample to determine how rotation periods in white dwarfs differ as a function of the overall white dwarf mass. We see in Figure~\ref{fig:rotation} suggestions of a trend of decreasing rotation period with increasing white dwarf mass. We also include on the right axis estimates for the ZAMS progenitor mass, calculated from the cluster-calibrated white dwarf initial-to-final-mass relation of \citet{2016ApJ...818...84C} for progenitor stars with masses below 4.0\,\msun.

A first-order linear fit to the white dwarf mass-rotation plot in the right panel of Figure~\ref{fig:rotation} yields an estimate of the final white dwarf rotation rate as a function of progenitor mass: $P_{\mathrm{rot}} = \big( (2.36\pm0.14) - M_{\mathrm{ZAMS}} \big) / (0.0024\pm0.0031)$, where $P_{\mathrm rot}$ is the final rotation period expressed in hours and $M_{\mathrm ZAMS}$ is the ZAMS progenitor mass in solar masses. This linear trend is not yet statistically significant; we are actively seeking more white dwarfs with masses exceeding 0.72\,\msun\ (which likely arise from $>3.0$\,\msun\ ZAMS progenitors) to improve estimates of this relationship.

\begin{figure}
\centering{\includegraphics[width=0.95\columnwidth]{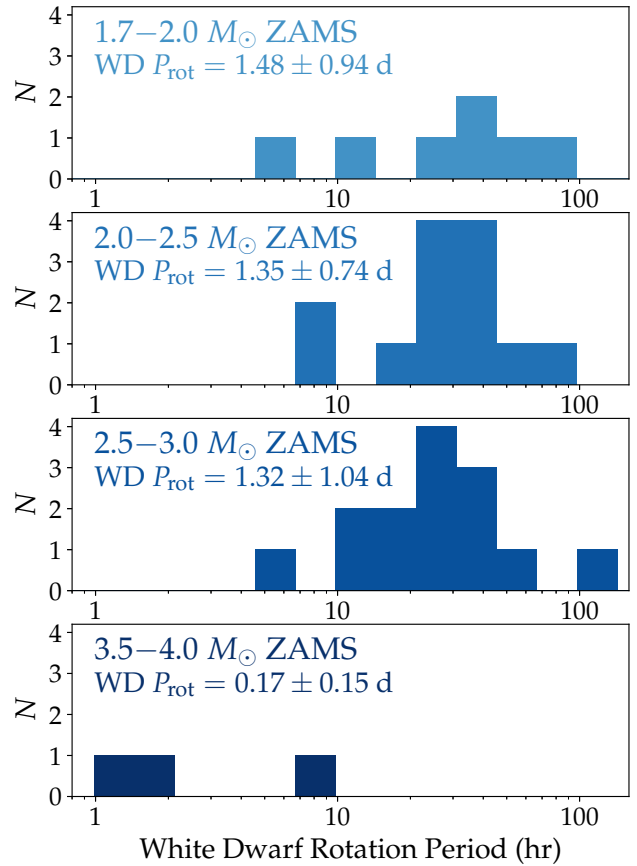}}
\caption{Rotation rates of pulsating, apparently isolated white dwarfs as a function of initial stellar mass, using the cluster-calibrated IFMR from \citet{2016ApJ...818...84C}. The 34 pulsating white dwarfs with masses from $0.51-0.73$\,\msun\ (which descended from roughly $1.7-3.0$\,\msun\ ZAMS stars) all have a mean rotation period of roughly 35 hr regardless of binning, but more massive white dwarfs ($>$$0.75$\,\msun, which descended from $>$$3.5$\,\msun\ ZAMS stars) appear to rotate systematically faster. \label{fig:rotbins}}
\end{figure}

The mass distribution of white dwarfs with rotation periods measured from pulsations is very similar to the mass distribution of field white dwarfs determined by \citet{2016MNRAS.461.2100T}. If we restrict our analysis to only white dwarfs within 1$\sigma$ of the mean mass and standard deviation of field white dwarfs (i.e., nearly 70\% of all field white dwarfs), we are left with 34 pulsating white dwarfs with masses from $0.51-0.73$\,\msun\ which have a mean rotation period of 35\,hr with a standard deviation of 28\,hr ($35\pm28$\,hr). The mean rotation period is essentially the same if we break the subsample into progenitor ZAMS masses spanning $1.7-2.0$\,\msun\ ($35\pm23$\,hr), $2.0-2.5$\,\msun\ ($32\pm18$\,hr), and $2.5-3.0$\,\msun\ ($32\pm25$\,hr). However, the small sample of three massive white dwarfs, which evolved from $3.5-4.0$\,\msun\ progenitors, appear to rotate significantly faster, at $4.0\pm3.5$\,hr. This is visualized in Figure~\ref{fig:rotbins}.

The original {\em Kepler} mission yielded exceptional insight into the rotational evolution of stars at most phases of stellar evolution \citep{2015AN....336..477A}, and now we can provide final boundary conditions on the question of internal angular momentum evolution in isolated stars. For example, {\em Kepler} data have been especially effective at illuminating differential rotation and angular momentum evolution in $1.0-2.0$\,\msun\ first-ascent red giants (e.g., \citealt{2012A&A...548A..10M}), where it was found that their cores are rotating roughly 10 times faster than their envelopes. Still, this is an order of magnitude slower than expected accounting for all hypothesized angular momentum transport processes (e.g., \citealt{2013A&A...549A..74M,2014ApJ...788...93C}). Moving further along the giant branch, {\em Kepler} has shown that $2.2-2.9$\,\msun\ core-helium-burning secondary clump giants do not have as much radial differential rotation and have cores rotating at roughly $30-100$\,days (e.g., \citealt{2015A&A...580A..96D}).

For decades we have known that white dwarfs have relatively slow rotation, given their surface rotation velocities usually do not exceed upper limits of 15\,\kms, corresponding to periods longer than several hours \citep{1998A26A...338..612K}. This is surprising: if we completely conserve the angular momentum of a 3.0\,\msun\ main-sequence star with an initial rotation period of 10\,hr, its white dwarf remnant would be rotating faster than a few minutes \citep{2004IAUS..215..561K}. We establish here that the majority of isolated descendants of $1.7-3.0$\,\msun\ ZAMS progenitors rotate at 1.5\,d, with a narrow dispersion between $0.5-2.2$\,d at the conclusion of their evolution. We note that we are sensitive to rotation periods longer than 15\,d given our frequency resolution, but the longest rotation period we measure is 109.1\,hr (roughly 4.5\,d) in EPIC\,228682478.

\section{Notes on Individual Objects}
\label{sec:individual}

\begin{figure}
\centering{\includegraphics[width=0.95\columnwidth]{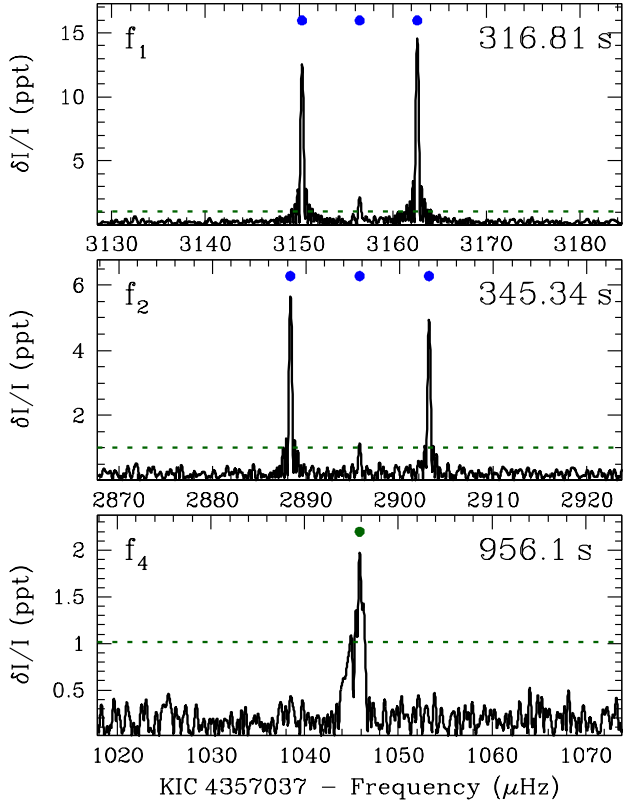}}
\caption{The highest-amplitude modes in KIC\,4357037 (KISJ1917+3927), $f_1$ and $f_2$ ($\ell=1$ modes marked with blue dots), illuminate the median $\ell=1$ splitting of 6.8\,\muhz, corresponding to a 22.0-hr rotation period. We could not identify the longest-period mode also present in this star at 956.11\,s (unidentified modes marked with green dots), which has a considerably broader line width than the shorter-period modes. \label{fig:4357037}}
\end{figure}
{\em KIC\,4357037 (KISJ1917+3927)}: With ``only'' 36.3\,d of nearly uninterrupted {\em Kepler} observations, this is one of our shortest datasets, but still the data reveal seven independent pulsations, five of which we identify as $\ell=1$ modes; we show in Figure~\ref{fig:4357037} the two highest-amplitude modes, $f_1$ and $f_2$. As noted in Section~\ref{sec:spectroscopy}, the atmospheric parameters for this object were misreported in \citet{2016MNRAS.457.2855G}, but are corrected here and find \teff\ $= 12{,}650$\,K. Most periods we observe are in line with such a hot DAV, with the identified $\ell=1$ modes spanning $214.40-396.70$\,s. However, there are several long-period modes that do not appear to be nonlinear difference frequencies, at 549.72\, and 956.11\,s. The latter mode at 956.11\,s (bottom panel of Figure~\ref{fig:4357037}) is significantly broader than the spectral window, with HWHM=0.318\,\muhz, suggesting it is an independent mode in the star, given its position in Figure~\ref{fig:periodhwhm}. It is at a much longer period than typically observed in such a hot DAV, posing an interesting question of how it is excited at the same time as other modes with three times shorter periods.

{\em KIC\,4552982 (KISJ1916+3938)}: This was the first DAV discovered in the original {\em Kepler} mission field and has the longest pseudo-continuous light curve of any pulsating white dwarf ever recorded, spanning more than 1.5\,yr. Our analysis here is focused on the HWHM of the modes present, and thus only includes roughly 3 months of data from Q12, to remain consistent with the frequency resolution of the other DAVs analyzed here. A more detailed light curve analysis, including discussion of the mean period spacing and results from Lorentzian fits to the full {\em Kepler} dataset, can be found in \citet{2015ApJ...809...14B}. The triplet at $f_1$ remains the only mode we have identified in this DAV; the large amplitude of this coherent mode at $f_1$ drives down the WMP of this DAV to 750.4\,s, which we report from the period list of \citet{2015ApJ...809...14B}. Our rotation rate differs slightly from that reported by \citet{2015ApJ...809...14B} since we adopt a lower assumed value for $C_{k,\ell}$.

\begin{figure}
\centering{\includegraphics[width=0.95\columnwidth]{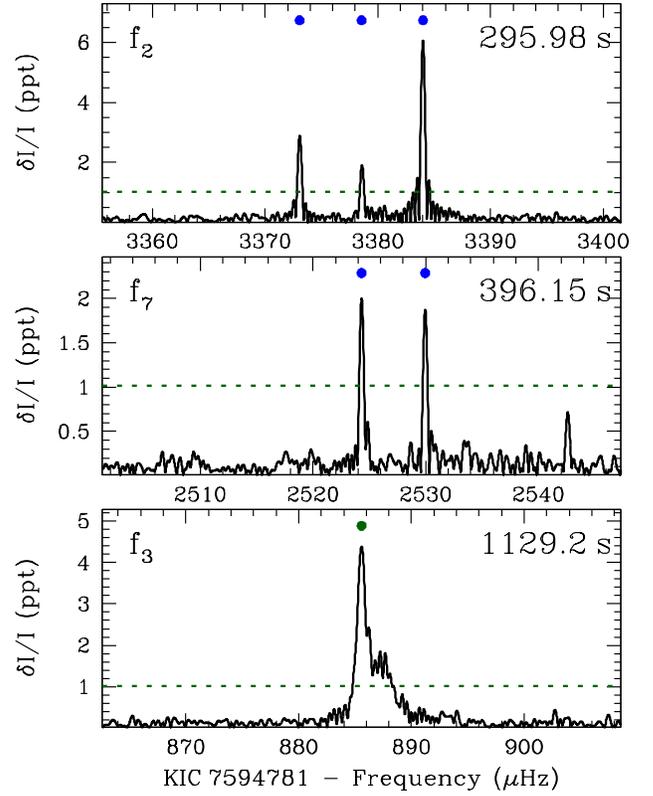}}
\caption{We show two modes in KIC\,7594781 (KISJ1908+4316) which share the median $\ell=1$ splitting of 5.5\,\muhz, suggesting a rotation period of 26.8\,hr. This DAV has a broadened long-period mode at $f_3$ (1129.19\,s) which appears in many of the nonlinear combination frequencies (Table~\ref{tab:pulsations}). \label{fig:7594781}}
\end{figure}
{\em KIC\,7594781 (KISJ1908+4316)}: This DAV has a relatively short dataset, with just 31.8\,d of observations in Q16.2. We identify three modes in the star as $\ell=1$ with a median splitting of 5.5\,\muhz, two of which are shown in Figure~\ref{fig:7594781}, suggesting a rotation period of 26.8\,hr. Such a rotation rate would yield $\ell=2$ splittings of roughly 8.6\,\muhz, so we suggest that $f_1$ may be an $\ell=2$ mode, although the data are not definitive. As with KIC\,4357037, there is a much longer-period, highly broadened mode, here at 1129.19\,s. Interestingly, this mode, $f_3$, combines with many other modes in the star to give rise to the most of the 14 nonlinear combination frequencies present. When $f_3$ combines with other modes its unique line width is reproduced; for this reason, although we could not identify any suitable combinations, we caution that $f_4$ may not be an independent mode in the star but could possibly be a combination frequency of $f_3$ in some way given the shape of $f_4$. Interestingly, $f_3$ is relatively stable in amplitude but changing rapidly in frequency. Its frequency in the first 5\,days of data is relatively stable at $885.243\pm0.057$\,\muhz\ ($7.08\pm0.32$\,ppt), but its power in the last 5\,days of {\em Kepler} data less than 27\,d later had almost completely moved to $888.285\pm0.067$\,\muhz\ ($6.14\pm0.32$\,ppt). There is still much to be learned about the causes of the short-term frequency, amplitude, and phase instability in white dwarf pulsations from further analysis of this {\em Kepler} data --- nonlinear mode coupling appears to be the only way to explain how white dwarf pulsations can change character so rapidly (e.g., \citealt{2016A&A...585A..22Z}).

\begin{figure}
\centering{\includegraphics[width=0.95\columnwidth]{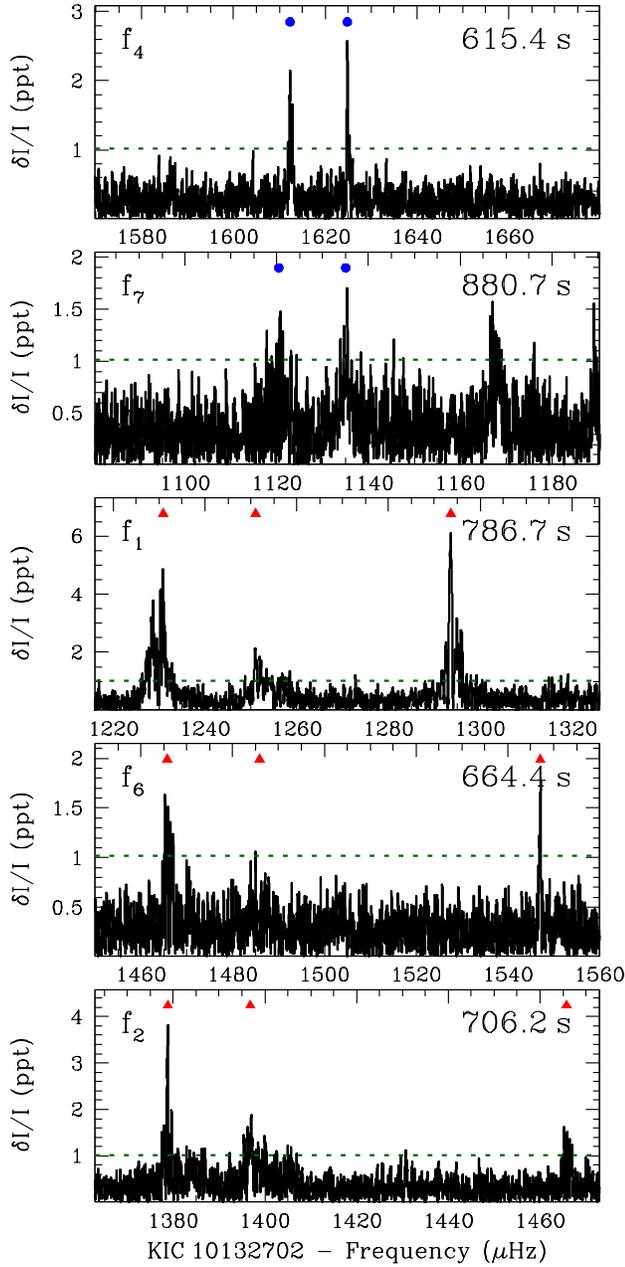}}
\caption{We show several components of $\ell=1$ modes (marked with blue dots) and $\ell=2$ modes (red triangles) in KIC\,10132702 (KISJ1913+4709). The top two panels highlight the two dipole modes, with a median $\ell=1$ splitting of 13.5\,\muhz. We also show what we have identified as three different quadrupole modes, with a median $\ell=2$ splitting of 21.3\,\muhz. Our tentative $m$ identification for these modes is detailed in Table~\ref{tab:pulsations}. Both sets of splittings suggest a rotation period of roughly 11.2\,hr. \label{fig:10132702}}
\end{figure}
{\em KIC\,10132702 (KISJ1913+4709)}: We are able to identify seven of the eight independent modes in this DAV, five of which are shown in Figure~\ref{fig:10132702}. The periods span roughly $461.1-977.6$\,s, and most have broad HWHM, with a median HWHM exceeding 0.35\,\muhz. Both our median $\ell=1$ splitting of 13.5\,\muhz\ and median $\ell=2$ splitting of 21.3\,\muhz\ suggest a roughly 11.2\,hr rotation period, confirming that our mode identifications are internally self-consistent.

{\em KIC\,11911480 (KISJ1920+5017)}: This hot DAV (also cataloged as KIC\,100003912) was thoroughly discussed in \citet{2014MNRAS.438.3086G}, who present an analysis of data from both Q12 and Q16, allowing for the identification of four of the five independent modes present. Here we only analyze data from Q16 but arrive at the same mode identifications.

\begin{figure}
\centering{\includegraphics[width=0.95\columnwidth]{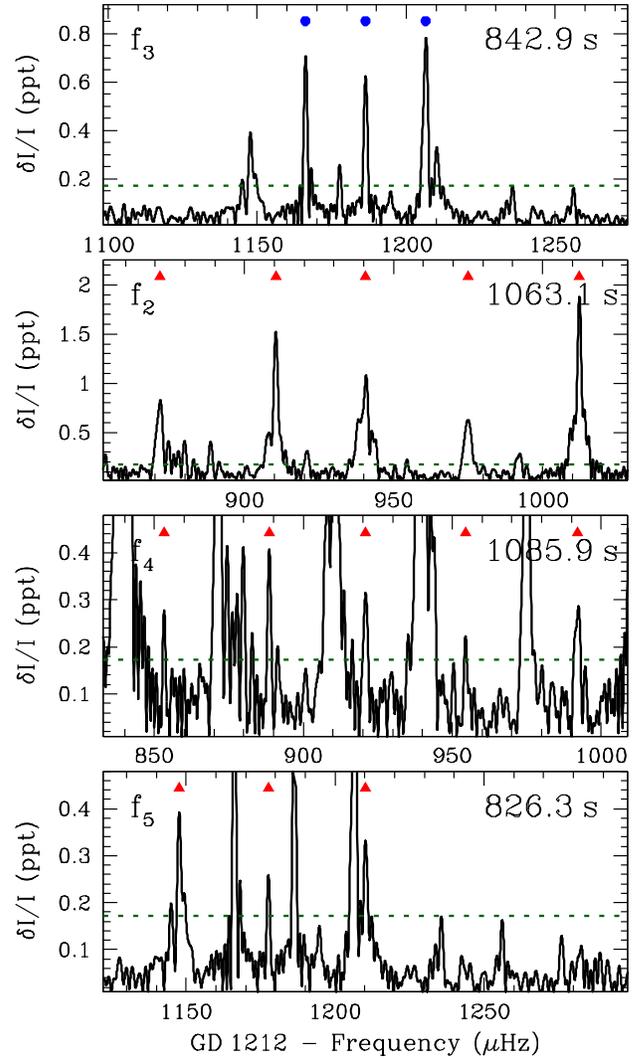}}
\caption{We show one $\ell=1$ mode (marked with blue dots) and three $\ell=2$ modes (red triangles) in GD\,1212. The dipole mode has an $\ell=1$ splitting of 21.3\,\muhz, and our three different quadrupole modes have a median $\ell=2$ splitting of 33.7\,\muhz. Our tentative $m$ identification for these modes is detailed in Table~\ref{tab:pulsations}. Both sets of splittings are self-consistent and reveal a rotation period of roughly 6.9\,hr. \label{fig:gd1212}}
\end{figure}
{\em EPIC\,60017836 (GD\,1212)}: We detailed our initial pulsation period list of GD\,1212 in \citet{2014ApJ...789...85H}, which was the first published result using data from the two-wheel-controlled {\em K2} mission. After an extensive analysis using only the final 9\,days of the {\em K2} engineering data, we have identified five of the modes present in the star, after concluding that the star rotates with a period of roughly 6.9\,hr. At the top of Figure~\ref{fig:gd1212} we show the only identified $\ell=1$ mode, $f_3$, which has a splitting of 21.3\,\muhz\ and is embedded within the $\ell=2$ mode $f_5$. Additionally, we find two consecutive quadrupole modes embedded within one another: $f_2$ and $f_4$; the median $\ell=2$ splitting is 33.7\,\muhz. There are additional modes, including many nonlinear combination frequencies, presented in \citet{2014ApJ...789...85H} from the full 11.5\,d of engineering data on GD\,1212, which were not significant from only this 9-d dataset. We note that {\em K2} has revisited GD\,1212 for more than 75\,d during Campaign 12.

\begin{figure}
\centering{\includegraphics[width=0.95\columnwidth]{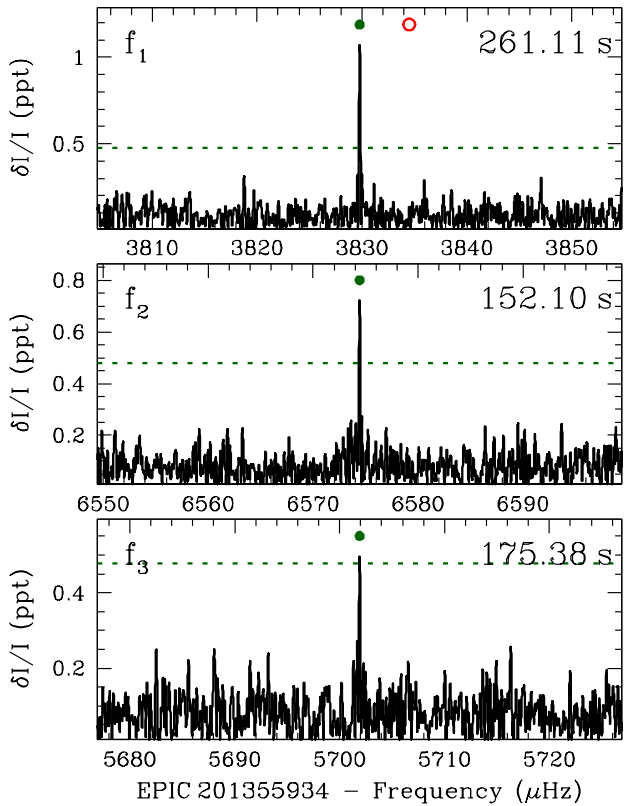}}
\caption{We show the three unidentified singlets in EPIC\,201355934 (SDSSJ1136$-$0136). We also mark with a red open circle the location of the only detected periodicity from ground-based discovery of pulsations in SDSSJ1136$-$0136, at 260.79\,s \citep{2010MNRAS.405.2561C}. \label{fig:201355934}}
\end{figure}
{\em EPIC\,201355934 (SDSSJ1136$-$0136)}: We observe just three significant peaks in this DAV, all shown in Figure~\ref{fig:201355934}. None of the peaks appear to be components of the same $k,\ell$ multiplet, which means we cannot estimate the rotation period of this white dwarf; this is possible if the star is seen roughly pole-on, such that we only observe the $m=0$ components. This DAV was known to pulsate before the launch of {\em Kepler}, but the only photometry is a discovery run yielding just one period at 260.79\,s (\citealt{2010MNRAS.405.2561C}, shown as open circle in Figure~\ref{fig:201355934}), nearest to $f_1$ in our {\em K2} observations.

\begin{figure}
\centering{\includegraphics[width=0.95\columnwidth]{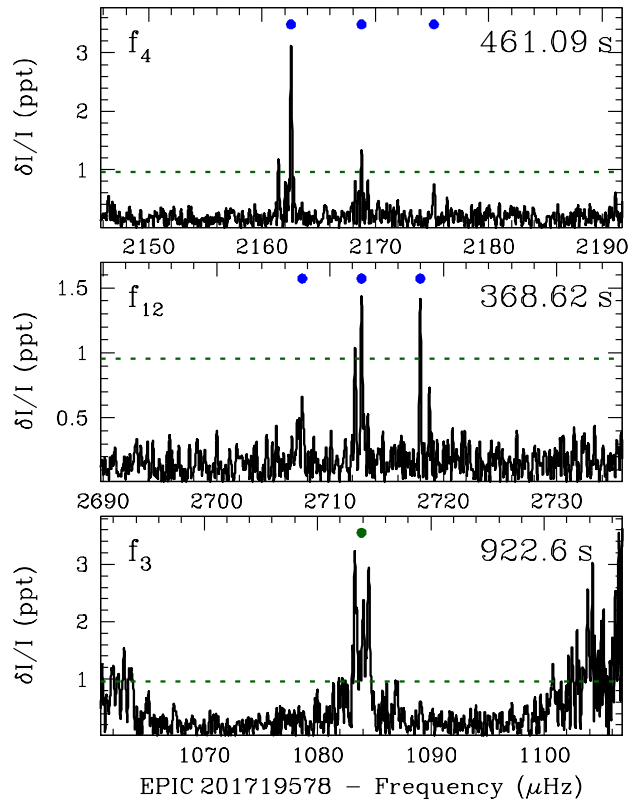}}
\caption{At top we show two identified dipole modes in EPIC\,201719578 (SDSSJ1122$+$0358); the median $\ell=1$ splitting of 5.5\,\muhz\ reveals a 26.8\,hr rotation period. The bottom panel shows a longer-period mode with a broader linewidth that we cannot definitively identify. \label{fig:201719578}}
\end{figure}
{\em EPIC\,201719578 (SDSSJ1122$+$0358)}: The periods in this cool DAV range from $367.9-1095.4$\,s, but only the three shortest-period modes are stable enough to be cleanly identified. We show in Figure~\ref{fig:201719578} two of the $\ell=1$ modes, which have a median splitting of 5.5\,\muhz, revealing this white dwarf's 26.8\,hr rotation period. The bottom panel of Figure~\ref{fig:201719578} shows the broader linewidth of a longer-period mode in the same star. It is unlikely that all 23 of the modes listed in Table~\ref{tab:pulsations} are independent modes. With an $\ell=1$ splitting of 5.5\,\muhz, we would expect $\ell=2$ splittings of roughly 8.6\,\muhz; therefore, for example, $f_3$ and $f_1$ could be the $m=-2$ and $m=+1$ components, respectively, of the same $\ell=2$ mode. There are therefore likely between $14-18$ independent modes in this DAV, depending on how many of the unidentified modes are components of the same multiplet. Previous ground-based photometry from the discovery of pulsations showed two significant periodicities, at 859.0\,s and 996.1\,s \citep{2004ApJ...607..982M}. We only see evidence for the mode $f_2$ at $861.4\pm4.0$\,s in this {\em K2} dataset (Table~\ref{tab:pulsations}).

{\em EPIC\,201730811 (SDSSJ1136$+$0409)}: The pulsations and light curve modeling of this DAV, which is the only pulsating white dwarf known in a post-common-envelope binary, were detailed in \citet{2015MNRAS.451.1701H}. We excluded this DAV from Figure~\ref{fig:rotation} since this white dwarf almost certainly interacted with its close dM companion, which currently orbits the DAV every 6.9\,hr. We have updated Table~\ref{tab:pulsations} to reflect our re-analysis of the pulsations: It appears that the two lowest-amplitude, longest-period oscillations are not independent modes but are likely difference frequencies (combination frequencies arising from nonlinear distortions in the atmosphere of the star). Specifically, it appears that the periodicity we called ``$f_6$'' in \citet{2015MNRAS.451.1701H} at 474.45\,s is within the $1\sigma$ uncertainties of the difference frequency $f_{5}-f_{4a}$. Additionally, what we called ``$f_7$'' at 395.91\,s is within the $1\sigma$ uncertainties of the difference frequency $f_{3b}-f_{1c}$. Omitting both modes does not substantially alter the asteroseismic analysis (A.~Bischoff-Kim, private communication).

{\em EPIC\,201802933 (SDSSJ1151$+$0525)}: We have shown the Fourier transforms of the five identified modes of this hot DAV, all of which are $\ell=1$ modes, in Figure~\ref{fig:ftc1} in Section~\ref{sec:photo}. Their median splitting is 4.7\,\muhz, which corresponds to a roughly 31.3\,hr rotation period. It is possible that the mode $f_6$ at 123.18\,s could suffer from the Nyquist ambiguity and instead be centered at 112.61\,s; in all cases here we have selected the Nyquist alias with the highest amplitude, but this is not a guarantee they are the true signals in the star \citep{2015MNRAS.453.2569M}. Unfortunately, the {\em K2} observations only cover a small portion of the {\em Kepler} spacecraft orbital motion, so we cannot search for incorrect Nyquist alias multiplet splittings at the orbital frequency of the spacecraft \citep{2013MNRAS.430.2986M}. The Nyquist ambiguity for $f_6$ could be settled with follow-up, ground-based photometry, but for now we presume the peaks centered around 123.18\,s are the correct aliases.

\begin{figure}
\centering{\includegraphics[width=0.95\columnwidth]{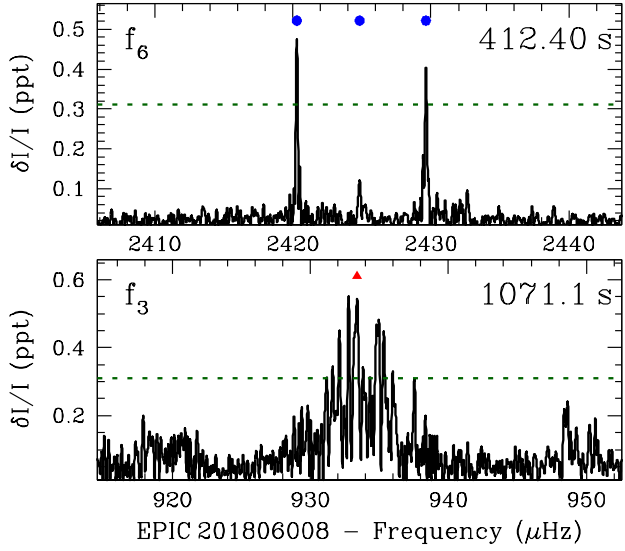}}
\caption{The top panel illuminates one of the identified dipole modes of EPIC\,201806008 (PG\,1149+057), which reveals the median $\ell=1$ splitting of 4.7\,\muhz. This is the brightest outbursting DAV \citep{2015ApJ...810L...5H} and has many modes with very broad bands of power; the Fourier transforms shown here include all {\em K2} data collected, including data taken during outbursts. The bottom panel details a longer-period mode with a much broader linewidth. \label{fig:201806008}}
\end{figure}
{\em EPIC\,201806008 (PG\,1149+057)}: We published an analysis of how outbursts affect the pulsations in this cool DAV in \citet{2015ApJ...810L...5H}, but did not thoroughly discuss the pulsations present in that letter. The Fourier transform we analyze here includes all {\em K2} data, including that taken in outburst. The shortest-period mode is most stable, and reveals an apparent $\ell=1$ splitting of 4.7\,\muhz\ (top panel of Figure~\ref{fig:201806008}), suggesting an overall rotation of 32.6\,hr and commensurate $\ell=2$ splittings of roughly 7.4\,\muhz. From this overall expectation, we tentatively identify six of the 14 independent modes present, most of which are highly broadened, an example of which is shown at the bottom panel of Figure~\ref{fig:201806008}. Excluding $f_6$ centered at 412.40\,s, all other periods in this cool DAV range from $833.7-1392.1$\,s.

\begin{figure}
\centering{\includegraphics[width=0.95\columnwidth]{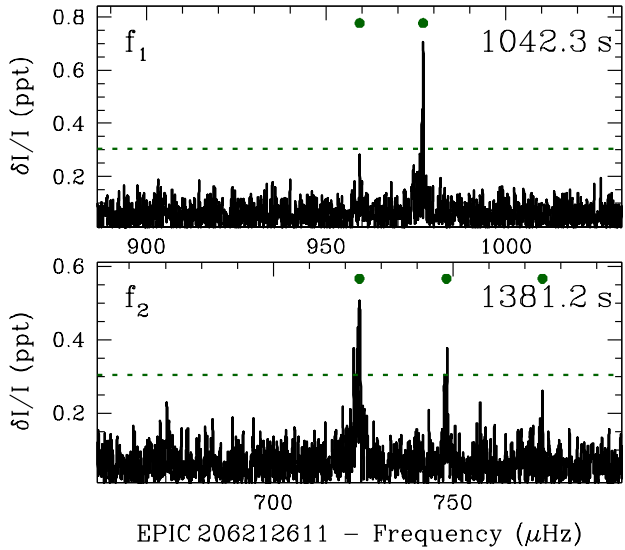}}
\caption{The two highest-amplitude sets of modes in EPIC\,206212611 (SDSSJ2220$-$0933) do not definitively reveal their identity. The two peaks in the top panel, which we refer to as $f_1$, suggest a possible $\ell=1$ splitting 17.7\,\muhz, and the three peaks we refer to as $f_2$ have a roughly 26\,\muhz\ splitting, which would both correspond to a rotation period of roughly 8.3\,hr. However, exclude this cool DAV from Section~\ref{sec:rotation} and Figure~\ref{fig:rotation}, since we have not definitively identified the pulsations present. \label{fig:206212611}}
\end{figure}
{\em EPIC\,206212611 (SDSSJ2220$-$0933)}: We observe just six significant peaks in this DAV, ranging from $1023.4-1381.2$\,s, and we show five of these six peaks in Figure~\ref{fig:206212611}. It is difficult to identify the pulsations with so few modes, but it is possible that $f_1$ has a $\sim$17.7\,\muhz\ splitting and $f_2$ has a $\sim$26\,\muhz\ splitting, both of which correspond to a rotation period of roughly 8.3\,hr if $f_1$ is an $\ell=1$ mode and $f_2$ is an $\ell=2$ mode. However, we have not included this rotation period in our analysis in Section~\ref{sec:rotation}, because we lack certainty on the mode identification.

\begin{figure}
\centering{\includegraphics[width=0.95\columnwidth]{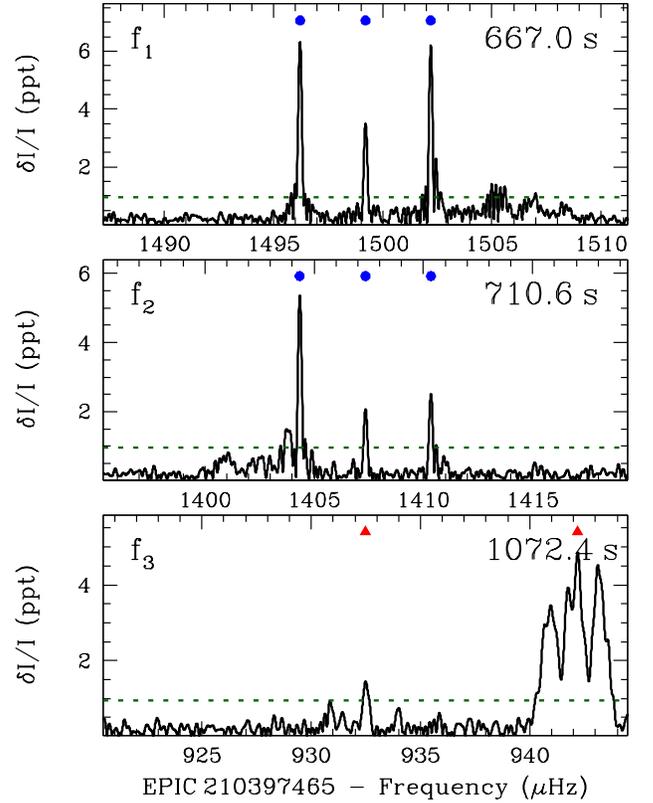}}
\caption{The top two panels show two $\ell=1$ modes in EPIC\,210397465 (SDSSJ0358$+$1324), which share the median splitting of 3.0\,\muhz, suggesting a rotation period of roughly 49.1\,hr. The bottom panel shows the likely $m=0$ and $m=+2$ components of an $\ell=2$ mode at much longer period, corroborating the rotation rate; the $m=+2$ component is significantly broadened, with a HWHM=0.963\,\muhz. Interestingly, both $f_1$ and $f_2$ show broad peaks outside the expected distribution for $\ell=1$ modes. \label{fig:210397465}}
\end{figure}
{\em EPIC\,210397465 (SDSSJ0358$+$1324)}: We detect at least 10 nonlinear combination frequencies of this DAV, most of which involve either $f_1$ or $f_2$, which we identify as $\ell=1$ modes and display in the top panels of Figure~\ref{fig:210397465}. The median $\ell=1$ splitting of 3.0\,\muhz\ suggests a rotation period of 49.1\,hr. We also observe three other multiplets which appear to have a median splitting of 4.7\,\muhz, exactly what we would expect for $\ell=2$ splittings for a 49.1-hr rotation period; one such multiplet, $f_3$, is shown in the bottom panel of Figure~\ref{fig:210397465}. We are therefore able to identify six of the 11 independent modes present. The highest-amplitude modes at 667.0\,s and 710.6\,s have relatively narrow line widths, with HWHM$<$0.06\,\muhz, commensurate with the spectral window of the observations. However, both $f_1$ and $f_2$ show significant, broad, unidentified power; for $f_1$ this power is at higher frequency than the $m=+1$ component, and in $f_2$ this power is at lower frequency than the $m=-1$ component (Figure~\ref{fig:210397465}). As with KIC\,7594781, further investigation of the frequency and amplitude stability of this DAV is warranted. We have used the C/O-core models of \citet{2001PASP..113..409F} to determine the mass of SDSSJ0358$+$1324, but this DAV is very near the boundary for He-core white dwarfs.

\begin{figure}
\centering{\includegraphics[width=0.95\columnwidth]{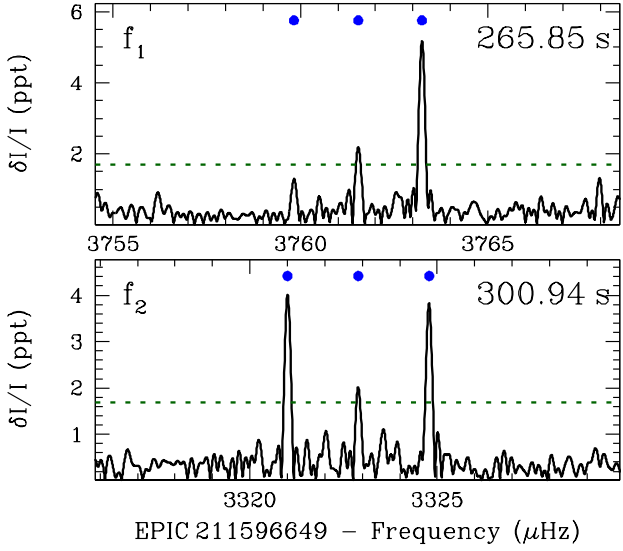}}
\caption{We show two modes in EPIC\,211596649 (SDSSJ0832$+$1429) which share the median $\ell=1$ splitting of 1.8\,\muhz, corresponding to a relatively long rotation period of roughly 3.4\,days.  \label{fig:211596649}}
\end{figure}
{\em EPIC\,211596649 (SDSSJ0832$+$1429)}: We observe just three independent modes in this DAV, two of which we identify as $\ell=1$ modes and display in Figure~\ref{fig:211596649}. The median splitting of these peaks is 1.8\,\muhz, corresponding to a rotation period of 81.8\,hr. It is possible that these three peaks are the $m=-2,0,+2$ components of $\ell=2$ modes (which would arise if the white dwarf rotates at roughly 5.3\,days). However, we prefer their identification as $\ell=1$ modes given the increased geometric cancellation of $\ell=2$ modes.

\begin{figure}
\centering{\includegraphics[width=0.95\columnwidth]{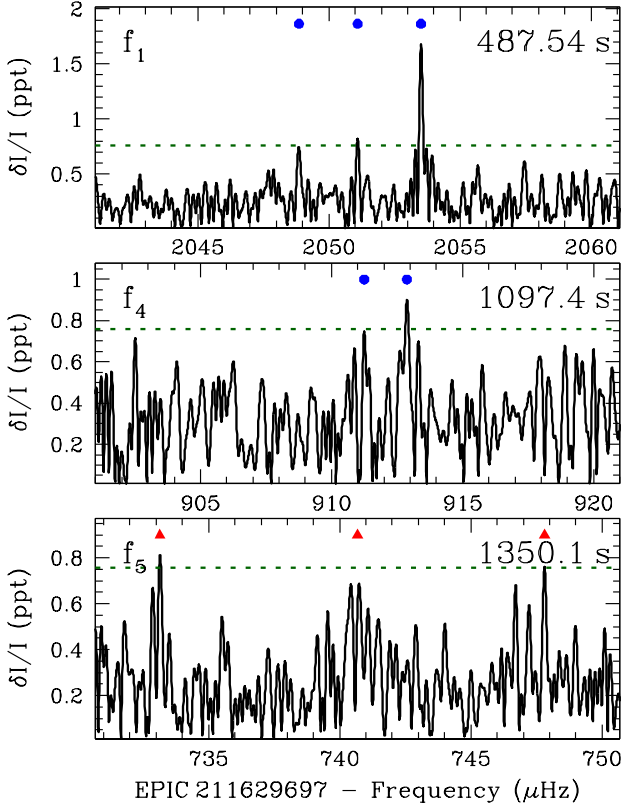}}
\caption{At top we show two dipole modes in EPIC\,211629697 (SDSSJ0840$+$1457), $f_1$ and $f_4$, which share the median $\ell=1$ splitting of 2.3\,\muhz. The bottom panel shows what we identify as the $m=-2,0,2$ components of an $\ell=2$ mode; the median $\ell=2$ splittings of 3.6\,\muhz\ corroborate the roughly 64.0\,hr rotation period in this outbursting DAV. \label{fig:211629697}}
\end{figure}
{\em EPIC\,211629697 (SDSSJ0840$+$1457)}: We have improved the extraction of this cool DAV from the discovery data described in \citet{2016ApJ...829...82B}, which analyzed the outburst behavior, and present analysis of the pulsations here. The shortest-period mode centered at 487.54\,s informs our analysis of the patterns present in the modes, and we have been able to identify seven of the eight independent modes observed in the star. The two top panels of Figure~\ref{fig:211629697} reveal the highest-amplitude $\ell=1$ modes, which have a median splitting of 2.3\,\muhz; this corresponds to a rotation period of 64.0\,hr. We would expect $\ell=2$ splittings of roughly 3.6\,\muhz\ from a 64-hr rotation period, which is exactly the splitting observed for the three peaks in $f_5$ shown in the bottom panel of Figure~\ref{fig:211629697}. Aside from the shorter-period $f_1$ centered at 487.54\,s, the other significant modes in this cool DAV range from $1095.5-1364.0$\,s.

{\em EPIC\,211914185 (SDSSJ0837$+$1856)}: We explored the pulsations of this massive DAV, including breaking the Nyquist ambiguity with follow-up, ground-based data, in \citet{2017ApJ...841L...2H}. Of white dwarfs with rotation rates measuring using asteroseismology, it remains the most massive and fastest rotating, likely the descendent of a roughly 4.0\,\msun\ ZAMS progenitor.

\begin{figure}
\centering{\includegraphics[width=0.95\columnwidth]{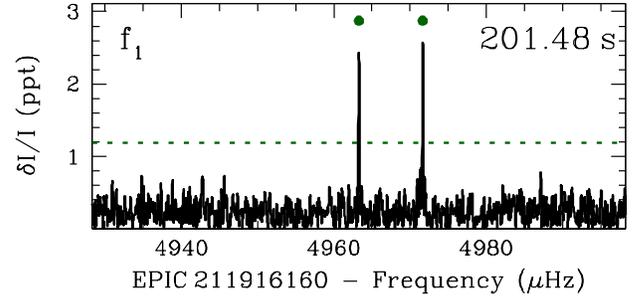}}
\caption{There are only two significant peaks in the Fourier transform of EPIC\,211916160 (SDSSJ0856$+$1858), separated by roughly 8.4\,\muhz, and so likely components of the same independent mode in the star. However, we lack sufficient information to identify this mode at 201.48\,s. \label{fig:211916160}}
\end{figure}
{\em EPIC\,211916160 (SDSSJ0856$+$1858)}: We have the fewest asteroseismic constraints on this DAV, since we only detect one independent mode, which we show in Figure~\ref{fig:211916160}. The two significant peaks we detect are separated by roughly 8.4\,\muhz, but we have no other constraints, and therefore cannot identify the spherical degree of this mode and do not include it in our rotation analysis in Section~\ref{sec:rotation}. The simplest explanation is that these are either the $m=0,+1$ or $m=-1,+1$ components of an $\ell=1$ mode, which would correspond to a rotation period of either 17.5\,hr or 8.8\,hr, respectively. However, for now we leave $f_1$ unidentified and do not include this DAV in our rotation analysis.

\begin{figure}
\centering{\includegraphics[width=0.95\columnwidth]{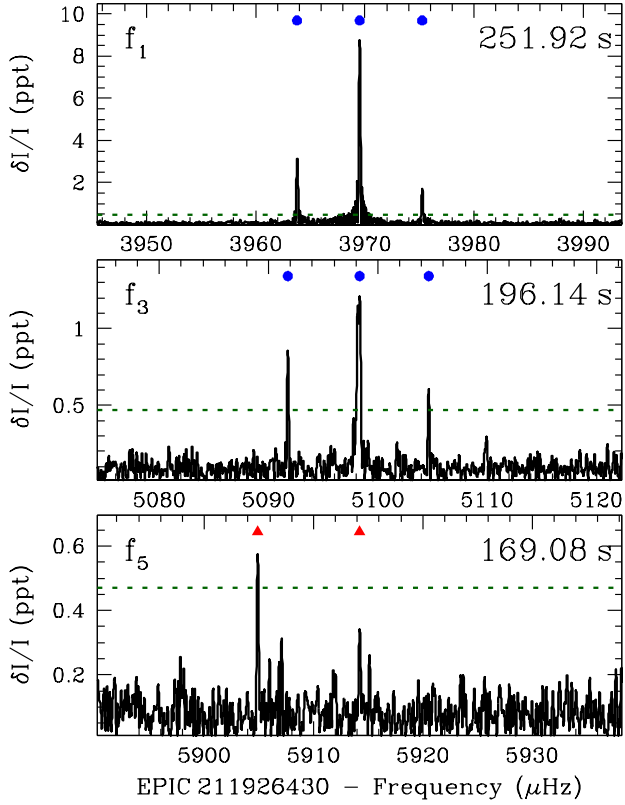}}
\caption{We show two modes in EPIC\,211926430 (SDSSJ0900$+$1907) in the top two panels which share the median $\ell=1$ splitting of 5.8\,\muhz, corresponding to a rotation period of 25.4\,hr. In the bottom panel we show the likely $m=-1,0$ components of an $\ell=2$ mode in the star; the median $\ell=2$ splitting of 9.1\,\muhz\ corroborates the 25.4\,hr rotation. \label{fig:211926430}}
\end{figure}
{\em EPIC\,211926430 (SDSSJ0900$+$1907)}: We are able to identify five of the six independent modes in this DAV. In the top two panels of Figure~\ref{fig:211926430} we show the highest-amplitude $\ell=1$ modes, which share the median splitting of 5.8\,\muhz, corresponding to a rotation period of 25.4\,hr. We also observe two $\ell=2$ modes with a median splitting of 9.1\,\muhz, fully consistent with the rotation period and confirming our mode identification. It is possible that the mode we observe centered at 119.51\,s is an incorrect Nyquist alias; the super-Nyquist alias of $f_6$ would instead be centered at roughly 115.96\,s, which could be ruled out or confirmed with follow-up photometry. With a spectroscopically determined temperature of $11{,}420$\,K, this is the coolest DAV in our sample with a WMP shorter than 400\,s; the periods we detect range from $119.51-299.58$\,s, with a WMP of 244.2\,s. However, it does not appear to be a significant outlier in Figure~\ref{fig:wmpteff}.

\begin{figure}
\centering{\includegraphics[width=0.95\columnwidth]{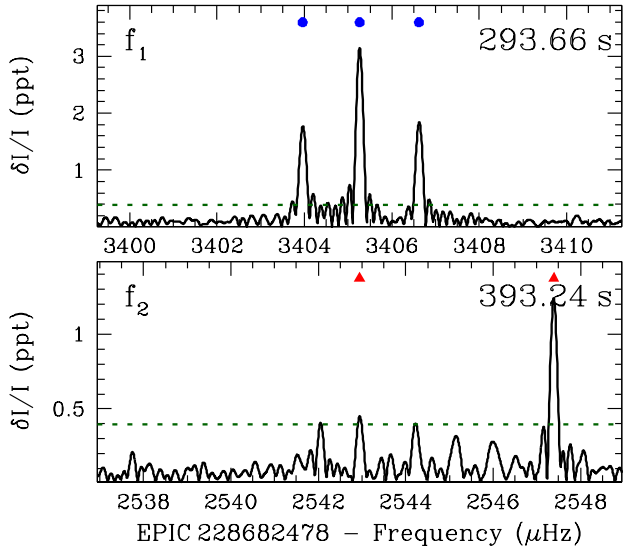}}
\caption{We show two modes in EPIC\,228682478 (SDSSJ0840$+$1320) which both suggest an overall rotation period of roughly 113.3\,hr: the $\ell=1$ mode $f_1$ and the the $\ell=2$ mode $f_2$. We only mark what appear to be the $m=0,+2$ components of $f_2$, but there are other nearly significant signals around this mode that we do not identify. \label{fig:228682478}}
\end{figure}
{\em EPIC\,228682478 (SDSSJ0840$+$1320)}: We have identified two of the three independent modes in this DAV, which we display in Figure~\ref{fig:228682478}. Both the splittings observed in the dipole mode $f_1$ (1.3\,\muhz) and the quadrupole mode $f_2$ (2.2\,\muhz) corroborate an overall rotation period of roughly 113.3\,hr (4.5\,days), the longest rotation period measured for any pulsating white dwarf. There are several other nearly significant signals around $f_2$ at $2544.231\pm0.014$\,\muhz\ ($0.39\pm0.07$\,ppt), $2542.053\pm0.014$\,\muhz\ ($0.39\pm0.07$\,ppt), $2545.152\pm0.016$\,\muhz\ ($0.36\pm0.07$\,ppt), and $2546.033\pm0.019$\,\muhz\ ($0.30\pm0.07$\,ppt). However, it is not clear if these represent an unfortunate confluence of noise (since none are formally significant), or if the represent multiplets from another independent mode around $f_2$, or if they are caused by some other means (such as the mysterious side peaks seen in $f_1$ and $f_2$ of EPIC\,210397465).

\begin{figure}
\centering{\includegraphics[width=0.95\columnwidth]{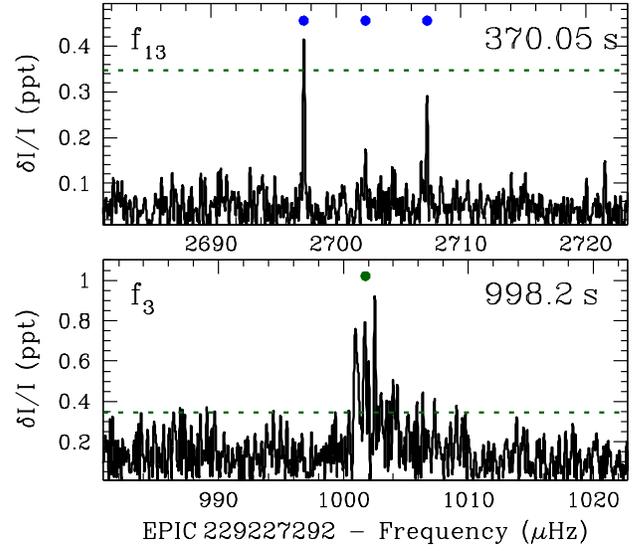}}
\caption{We show two modes in EPIC\,229227292 (ATLASJ1342$-$0735), including the $\ell=1$ mode $f_{13}$ which shares the median $\ell=1$ splitting of 5.0\,\muhz. Most other longer-period modes in this outbursting DAV have considerably broadened linewidths, making mode identification more difficult. \label{fig:229227292}}
\end{figure}
{\em EPIC\,229227292 (ATLASJ1342$-$0735)}: We show two modes in this outbursting DAV in Figure~\ref{fig:229227292}; the outburst characteristics were described in \citet{2016ApJ...829...82B}. We only identify three of the 18 independent modes in this rich pulsator, guided by the two short-period dipole modes at 370.05\,s and 288.91\,s, which share the median $\ell=1$ splitting of 5.0\,\muhz. Aside from those two short-period modes, the additional pulsations range from $514.1-1323.6$\,s, many with broad bands of power, such as $f_3$ shown in Figure~\ref{fig:229227292}.

\begin{figure}
\centering{\includegraphics[width=0.95\columnwidth]{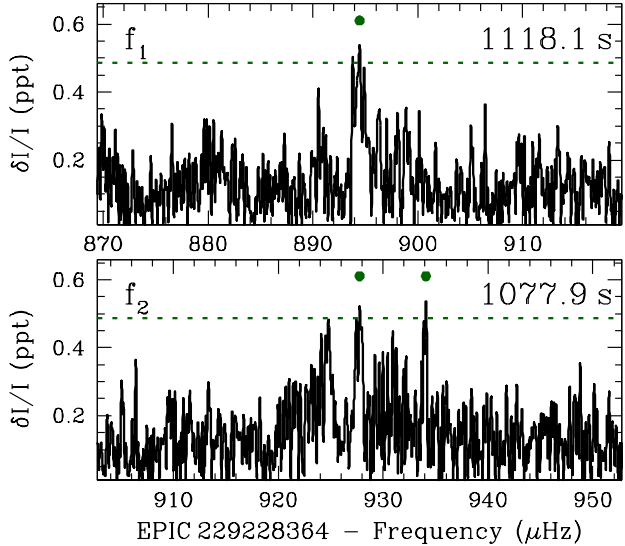}}
\caption{We show two modes in EPIC\,229228364 (SSSJ1918$-$2621) which are broadened and have low amplitude. We see two bands of power in $f_2$ separated by roughly 6.3\,\muhz, but do not have enough additional information to establish mode identification for the three independent modes present in this outbursting DAV. \label{fig:229228364}}
\end{figure}
{\em EPIC\,229228364 (SSSJ1918$-$2621)}: We detect at least six outbursts in this cool DAV, including a very long recurrence time between outbursts of at least 9.5 days (characterization of those outbursts will be presented in a forthcoming publication, as previewed in \citealt{2017ASPC..509..303B}). This DAV has the fewest pulsation modes of any outbursting DAV, with just three significant independent modes from $1070.6-1204.8$\,s (at $K_p$=17.9\,mag, this is not simply caused by low S/N). Only one of these three modes appears to be a multiplet, as seen in $f_2$ in the bottom panel of Figure~\ref{fig:229228364}; these modes are separated by 6.3\,\muhz. However, with just one incomplete multiplet, we cannot use it to inform our mode identification or rotation analysis.

\begin{figure}
\centering{\includegraphics[width=0.95\columnwidth]{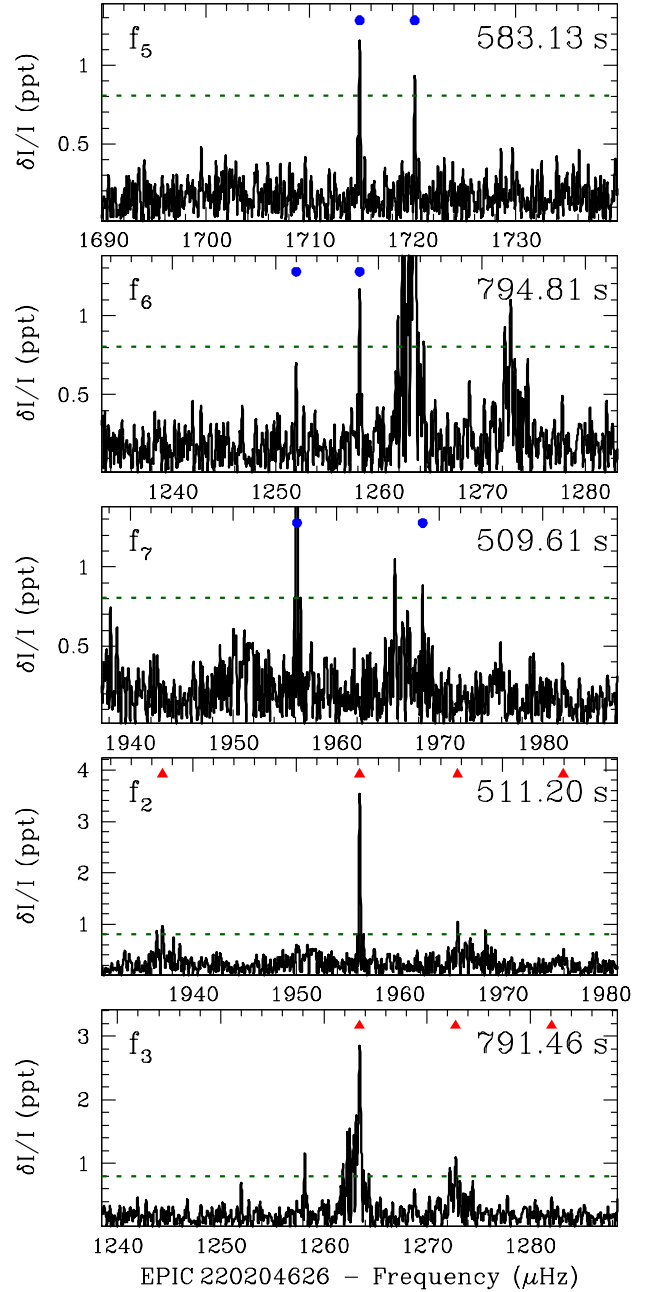}}
\caption{We show five modes in EPIC\,220204626 (SDSSJ0111$+$0009), some of which are closely overlapping, so we only mark the components of the respective multiplets with blue dots for the $\ell=1$ modes and red triangles for the $\ell=2$ modes. The splittings for all identified modes are consistent with an overall rotation period of roughly 24.3\,hr.  \label{fig:220204626}}
\end{figure}
{\em EPIC\,220204626 (SDSSJ0111$+$0009)}: This DAV has a line-of-sight M2 companion, but from the 37 SDSS subspectra we see no evidence of radial-velocity variability, so this is most likely a wide binary \citep{2016MNRAS.458.3808R}; this is corroborated by the lack of any low-frequency variability in the {\em K2} data. The 2.8-hr discovery light curve obtained on 2010~December~8 \citep{2015MNRAS.447..691P} found pulsations at periods of 631.6\,s (28\,ppt in $g'$), 583.2\,s (16.3\,ppt), 510.2\,s (18.9\,ppt), and 366.5\,s (9.1\,ppt). We see the three highest-amplitude of these periods in our {\em K2} data, as $f_4$, $f_5$, and $f_2$, respectively. Our extended light curve allows us to identify five of the seven independent pulsations present, which we show in Figure~\ref{fig:220204626}. The periods in this DAV range from $506.09-798.73$\,s, for a WMP of 642.7\,s. The median $\ell=1$ splitting of 6.1\,\muhz\ and median $\ell=2$ splitting of 9.5\,\muhz\ are both consistent with a 24.3\,hr rotation period. Many of the $\ell=1$ and $\ell=2$ modes are close together, and it appears that the $m=-1$ component of the dipole mode $f_7$ is very nearly overlapping in frequency with the $m=0$ component of the quadrupole mode $f_2$, shown one after the other in Figure~\ref{fig:220204626}. Notably the two $\ell=2$ modes have much higher amplitudes than the three identified $\ell=1$ modes, which is the opposite of what we would expect from geometric cancellation.

\begin{figure}
\centering{\includegraphics[width=0.95\columnwidth]{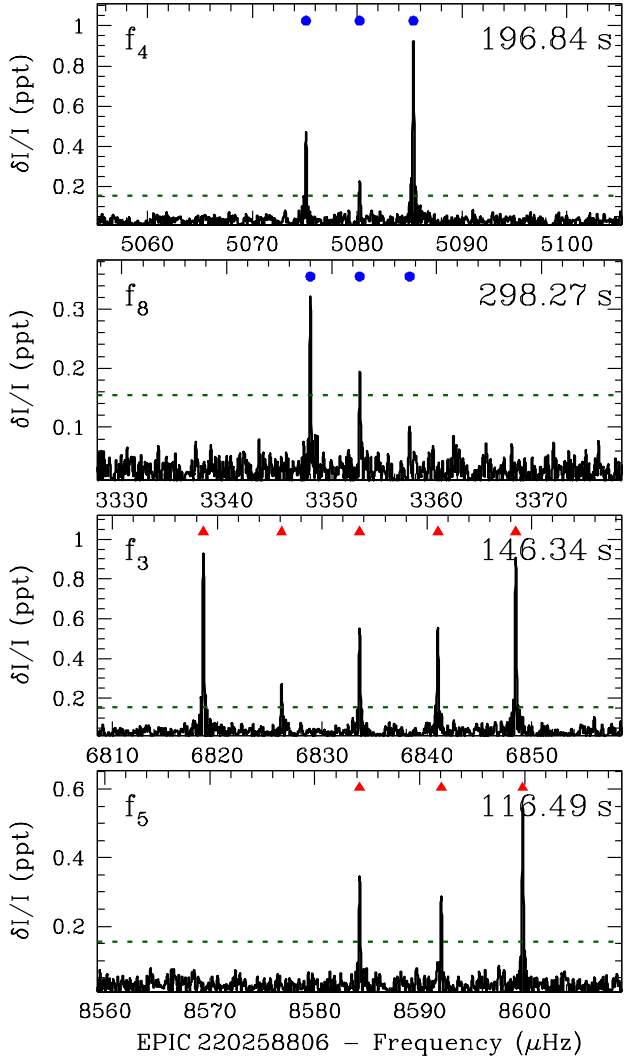}}
\caption{The hot DAV EPIC\,220258806 (SDSSJ0106$+$0145) provides an exceptional case of mode identification via rotational splittings. We plot the two $\ell=1$ modes, which share a median splitting of 4.9\,\muhz, as well as the two highest-amplitude $\ell=2$ modes (red triangles), which share a median splitting of 7.7\,\muhz. Both sets of splittings corroborate an overall rotation period of 30.0\,hr. \label{fig:220258806}}
\end{figure}
{\em EPIC\,220258806 (SDSSJ0106$+$0145)}: This hot DAV yields a textbook case of mode identification via rotational splittings, as shown in Figure~\ref{fig:220258806}. We are able to identify five of the 11 independent periods present in the star. The two dipole modes have a median splitting of 4.9\,\muhz, and the three quadrupole modes have a median splitting of 7.7\,\muhz, which both independently reveal the rotation period of 30.0\,hr. However, the two highest-amplitude modes appear as singlets and are not identified. There are two sets of signals very near the Nyquist frequency: the three peaks of $f_5$ have much higher amplitudes at the superNyquist position, so we use that Nyquist alias as the correct one in the star, whereas $f_{11}$ appears highest below the Nyquist frequency. Using three nights of time-series photometry from SOAR we confirm our selection of these aliases: we confirm that $f_5$ is superNyquist with periods centered at 116.49\,s, and that $f_{11}$ is just below the Nyquist frequency, with a period of 119.88\,s.

\begin{figure}
\centering{\includegraphics[width=0.95\columnwidth]{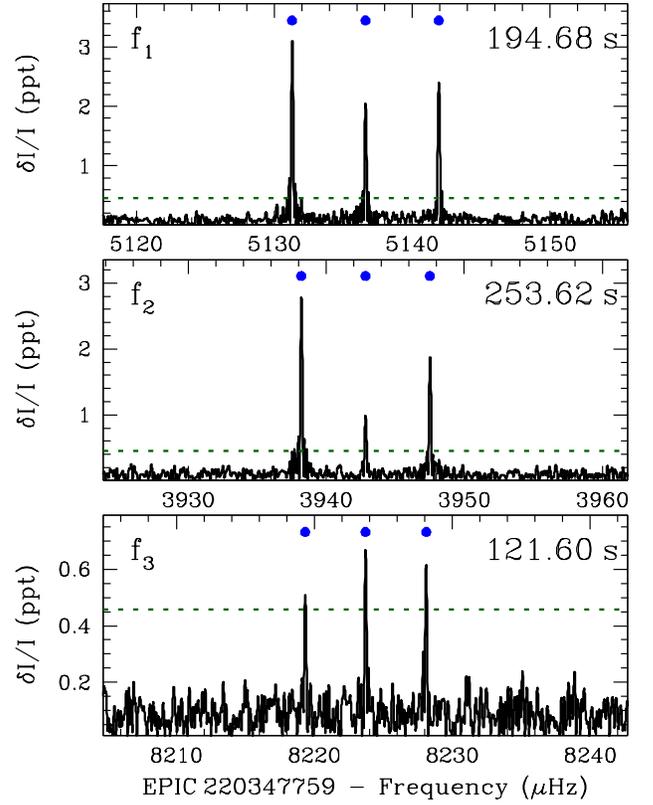}}
\caption{We show three modes in EPIC\,220347759 (SDSSJ0051$+$0339), all of which share roughly the median $\ell=1$ splitting of 4.65\,\muhz. \label{fig:220347759}}
\end{figure}
{\em EPIC\,220347759 (SDSSJ0051$+$0339)}: We show the three highest amplitude modes in Figure~\ref{fig:220347759}, all of which we identify as dipole modes with a median $\ell=1$ splitting of 4.65\,\muhz. There is some crowding in the field, with a brighter ($\Delta K_p<1.1$\,mag) star less than 13\arcsec\ from the target. 

\begin{figure}
\centering{\includegraphics[width=0.95\columnwidth]{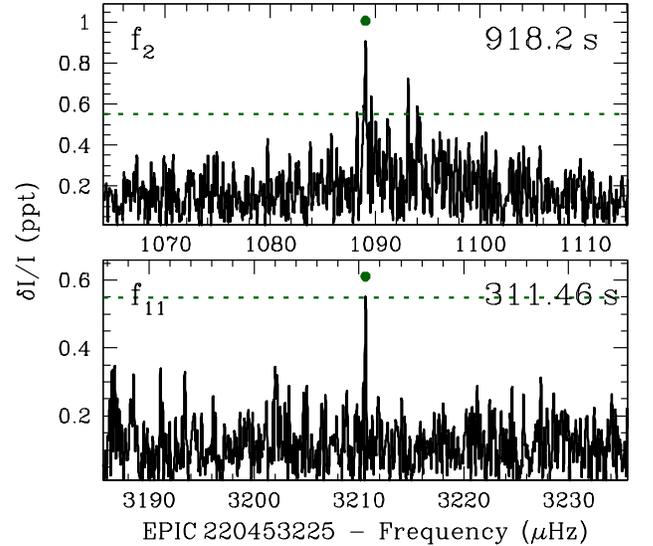}}
\caption{We show two modes in EPIC\,220453225 (SDSSJ0045$+$0544), neither of which we have been able to identify. This is the sixth outbursting DAV detected by the {\em Kepler} spacecraft \citep{2017ASPC..509..303B}. \label{fig:220453225}}
\end{figure}
{\em EPIC\,220453225 (SDSSJ0045$+$0544)}: We have detected at least 15 outbursts in this cool DAV, which we will characterize in a forthcoming publication. We show in Figure~\ref{fig:220453225} two different pulsations in the star: $f_2$ is representative of the longer-period pulsations present from $670.9-1391.5$\,s, and $f_{11}$ at 311.46\,s is a short-period mode also observed. We have not been able to solve any of the mode identifications for the 16 pulsations detected.

\begin{figure}
\centering{\includegraphics[width=0.95\columnwidth]{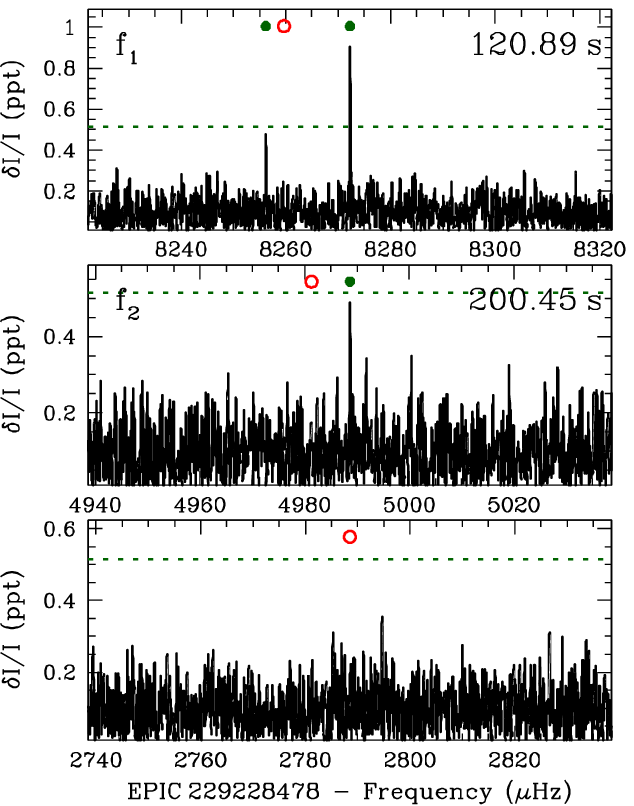}}
\caption{We show three frequency regions in the amplitude spectrum of EPIC\,229228478 (SDSSJ0122$+$0030). The open red circles denote the three periodicites detected from discovery photometry published by \citet{2010MNRAS.405.2561C}; we have discovered modes corresponding to the two highest-amplitude peaks from that work, but do not corroborate their mode at 358.6\,s. We have not identified any of the modes present in this hot DAV. \label{fig:229228478}}
\end{figure}
{\em EPIC\,229228478 (SDSSJ0122$+$0030)}: A short run from discovery ground-based photometry published by \citet{2010MNRAS.405.2561C} revealed three periodicities in this DAV, at 121.1\,s (1.5\,ppt), 200.8\,s (1.3\,ppt), and 358.6\,s (right at the significance threshold of 1.2\,ppt). We have detected two of these three ground-based modes, as shown in Figure~\ref{fig:229228478}, and show that the mode near 121.1\,s is likely a multiplet split by roughly 16.1\,\muhz. However, with just two multiplet components, we cannot identify any modes. If we assume that $f_1$ is an $\ell=1$ mode split by either 16.1\,\muhz\ or 8.0\,\muhz, that would correspond to an overall rotation rate of either 9.1\,hr or 18.4\,hr, respectively. However, since we lack firm mode identifications, we do not include this DAV in our rotation analysis.

\begin{figure}
\centering{\includegraphics[width=0.95\columnwidth]{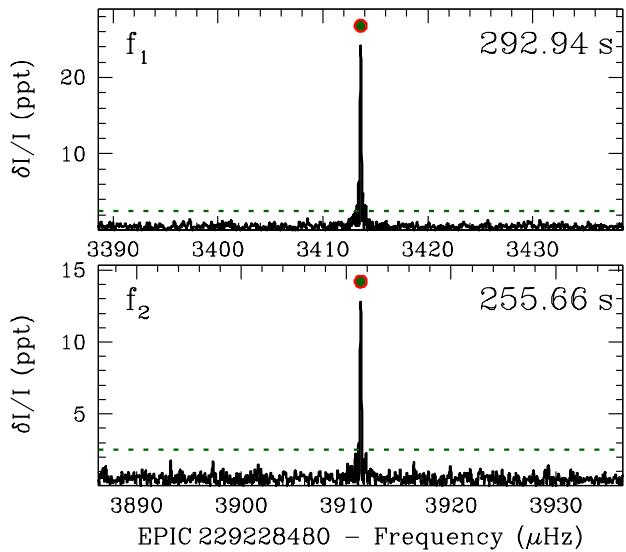}}
\caption{The hot DAV EPIC\,229228480 (SDSSJ0111$+$0018) shows just two independent modes, which appear as singlets. The periods determined from the {\em K2} match exactly the periods determined from the ground-based discovery \citep{2004ApJ...607..982M} and follow-up \citep{2013ApJ...766...42H} photometry. \label{fig:229228480}}
\end{figure}
{\em EPIC\,229228480 (SDSSJ0111$+$0018)}: This hot DAV has been monitored with ground-based photometry for more than a decade, and its pulsations reveal a rate of period change inconsistent with secular cooling \citep{2013ApJ...766...42H}. As with the long ground-based datasets, with {\em K2} we see only two independent modes that appear as singlets, shown in Figure~\ref{fig:229228480}, as well as nonlinear combination frequencies of those two modes. It is possible that this DAV is seen pole-on, such that only the $m=0$ components are visible; without multiplets, we cannot constrain mode identification.

\section{Discussion and Conclusions}
\label{sec:conclusion}

We present here the first bulk reduction and analysis of data collected on 27 pulsating DA white dwarfs observed by the {\em Kepler} space telescope up to {\em K2} Campaign 8. This long-duration, high-duty cycle photometery provides a way to securely resolve and identify hundreds of modes in DAVs without aliasing problems. The total baseline of data collected by {\em Kepler} and {\em K2} on DAVs already exceeds the sum total of ground-based optical photometry acquired since their identification as a class by \citet{1979ApJ...229..203M}.

The datasets analyzed here represent an extensive series of observations. The centerpiece is more than 2.75 million minute-cadence exposures collected during both the original {\em Kepler} mission as well as {\em K2}, yielding more than $45{,}000$\,hr (1875\,d) of space-based photometry of 27 pulsating white dwarfs. We have also obtained more than 140 new spectra collected over 15 separate nights with the 4.1-m SOAR telescope in order to characterize the atmospheric parameters of the entire sample, yielding external estimates of the white dwarf effective temperature and surface gravity, and thus overall white dwarf mass.

Our target selection is relatively unbiased; given the sparseness of white dwarfs in the {\em K2} observed so far, we have proposed many candidates for short-cadence photometry, most based on either serendipitous SDSS spectra or appropriate ($u$$-$$g$, $g$$-$$r$) colors. We therefore find pulsating white dwarfs throughout all regions of the DAV instability strip, spanning the stable, short-period hot DAVs at the hot edge of the strip all the way down to the coolest DAVs with the longest-period pulsations.

Our efforts in this paper have narrowly focused on the observational insights revealed from {\em Kepler} and now {\em K2}, leaving asteroseismic analysis of these stars to future publications. However, we have noticed some exceptional patterns, just from direct analysis of these unique observations that are free of diurnal aliasing that complicate ground-based observations.

We have discovered a significant dichotomy of mode linewidths, often within the same star, which appear almost completely related to the period of pulsation variability. Almost all pulsations with periods less than 800\,s appear to show relatively narrow linewidths; Lorentzian fits find their HWHM is comparable to the spectral window. However, most modes exceeding 800\,s show highly broadened linewidths, many of which resemble stochastic oscillators like the Sun. The measured HWHM of Lorentzian fits to these long-period modes correspond to $e$-folding timescale of order days to weeks, in line with expected growth (and damping) rates for high-radial-order modes. We are exploring a consistent physical model to explain these broad linewidths (Montgomery et al., in prep.).

We have also been able to identify the spherical degree ($\ell$) of roughly 40\% of the 154 independent modes present in these 27 DAVs, based on frequency patterns caused by rotational splittings. We are therefore able to measure internal rotation (still confined to at least the outer 1\% of the star) for 20 of the 27 DAVs observed so far with the {\em Kepler} spacecraft. 

Although we have not performed any asteroseismic fits on our sample (and many of these DAVs show just a few independent modes), we are able to use our follow-up spectroscopic observations to constrain, for the first time in a systematic way, white dwarf rotation as a function of mass. We find that white dwarfs with masses between $0.51-0.74$\,\msun\ (within 1$\sigma$ of the field white dwarf mass distribution) have a mean rotation period of 35\,hr and a standard deviation of 28\,hr. Using cluster-calibrated initial-to-final mass relations, such white dwarfs evolved from $1.7-3.0$\,\msun\ ZAMS progenitors. Therefore, we are putting narrowing boundary conditions on the endpoints of angular momentum evolution in the range of masses probed deeply by the {\em Kepler} spacecraft in earlier stages of stellar evolution. Given the slow rotation of apparently isolated white dwarfs, it thus appears from {\em Kepler} that most internal angular momentum is lost on the first-ascent giant branch. We hope that bulk white dwarf rotation rates as a function of mass can shed further light on the unknown angular momentum transport mechanism coupling red giant cores to their envelopes, especially the timescale of the coupling required to match internal rotation at all late stages of stellar evolution (e.g., \citealt{2013ApJ...775L...1T,2014ApJ...788...93C}).

{\em Kepler} photometry has shed insight into what appear to be at least five stages of DAV evolution, which we summarize in Figure~\ref{fig:colorfts}. Pulsation amplitudes start relatively low at the blue edge of the DAV instability strip, grow as the star cools, and eventually reach very high amplitudes when the star reaches the middle of the instability strip. We generally confirm the observational trend that cooler DAVs have longer pulsation periods, predicted from the theory that as white dwarfs cool their convection zones deepen and thus have a longer thermal adjustment timescale.

We also observe from {\em Kepler} and {\em K2} observations that nonlinear combination frequencies are most often visible for stars nearer the middle of the instability strip. We observe these nonlinear artifacts in only eight of the 27 DAVs here. Six of the eight with nonlinear combination frequencies cluster in a relatively narrow region of the instability strip, with a WMP ranging from $248.4-358.1$\,s and spectroscopic temperatures ranging from $12{,}750-11{,}880$\,K. However, we do not observe nonlinear combination frequencies in any of the six outbursting DAVs so far. There are just two relatively cool DAVs showing combination frequencies: GD\,1212 ($11{,}280$\,K, WMP=1019.1\,s, with combinations detailed in \citealt{2014ApJ...789...85H}) and EPIC\,210397465 ($11{,}520$\,K, WMP=841.0\,s). Although we are not as sensitive to the highest-frequency nonlinear combinations given that the Nyquist frequency falls at roughly 8496.18\,\muhz, we have shown from some of our DAVs that we can detect super-Nyquist signals. Thus, we have confidence in our assertion that it appears that, for most cases, observed nonlinear combination frequencies are restricted to a relatively narrow range of the instability strip, seen most often for the highest-amplitude DAVs in the middle of the instability strip.

{\em K2} continues to observe new fields along the ecliptic until it runs out of fuel, which is projected to occur at some point between Campaigns $17-19$. With continued proposal success, we expect to observe more than 150 additional candidate pulsating white dwarfs from Campaign 9 until the end of the mission, leaving a large legacy dataset of white dwarf variability. We venture to maintain this and future DAV datasets online at {\em k2wd.org}, where we have collected the raw and reduced {\em Kepler} lightcurves as well as the raw and reduced spectroscopy for the community to directly explore and (re-)analyze.

\acknowledgments


Support for this work was provided by NASA through Hubble Fellowship grant \#HST-HF2-51357.001-A, awarded by the Space Telescope Science Institute, which is operated by the Association of Universities for Research in Astronomy, Incorporated, under NASA contract NAS5-26555;
NASA {\em K2} Cycle 4 Grant NNX17AE92G;
NASA {\em K2} Cycle 2 Grant NNX16AE54G to Iowa State University;
NSF grants AST-1413001 and AST-1312983;
the European Research Council under the European Union's Seventh Framework Programme (FP/2007-2013) / ERC Grant Agreement n. 320964 (WDTracer);
the European Union's Horizon 2020 Research and Innovation Programme / ERC Grant Agreement n. 677706 (WD3D); and 
the Science and Technology Research Council (ST/P000495/1).

Based on observations obtained at the Southern Astrophysical Research (SOAR) telescope, which is a joint project of the Minist\'{e}rio da Ci\^{e}ncia, Tecnologia, e Inova\c{c}\~{a}o da Rep\'{u}blica Federativa do Brasil, the U.S. National Optical Astronomy Observatory, the University of North Carolina at Chapel Hill, and Michigan State University.

{\it Facilities:} {\em Kepler}, {\em K2}, SOAR, SDSS


\clearpage
\LongTables
\begin{landscape}

\begin{deluxetable}{lrccccccccccccccc}
\tabletypesize{\scriptsize}
\tablecolumns{17}
\tablewidth{0pc}
\tablecaption{Pulsation information from linear least-squares (LLS, Section~\ref{sec:photo}) and Lorentzian (Lor., Section~\ref{sec:hwhm}) fits for the first 27 DAVs observed by {\em Kepler} and {\em K2}. \label{tab:pulsations}}
\tablehead{
        \colhead{KIC/EPIC} &
        \colhead{Mode} &        
        \colhead{LLS Per.} &
        \colhead{$\Delta$Per.} &
        \colhead{LLS Freq.} &
        \colhead{$\Delta$Freq.} &
        \colhead{Amp.} &  
        \colhead{$\Delta$Amp.} &
        \colhead{Phase} & 
        \colhead{$\Delta$Phase} &
        \colhead{Lor. Per.} &
        \colhead{$\Delta$Per.} &   
        \colhead{HWHM} &  
        \colhead{Lor. Amp.} &  
        \colhead{Splitting} & 
        \colhead{$\ell$} &
        \colhead{$m$} \\
 &  & (s) & (s) & (\muhz) & (\muhz) & (ppt) & (ppt) & (rad.) & (rad.) & (s) & (s) & (\muhz) & (ppt$^2$) & (\muhz) & &  }
\startdata
4357037 & $f_{1a}$ & 317.42718 & 0.00022 & 3150.3288 & 0.0021 & 12.41 & 0.15 & 0.6177 & 0.0019 & 317.427 & 0.011 & 0.108 & 172.777 & \nodata & 1 & $-$1 \\
4357037 & $f_{1b}$ & 316.8081 & 0.0013 & 3156.485 & 0.013 & 1.98 & 0.15 & 0.299 & 0.012 & 316.808 & 0.011 & 0.105 & 5.051 & 6.156 & 1 & 0 \\
4357037 & $f_{1c}$ & 316.19259 & 0.00018 & 3162.6294 & 0.0018 & 14.39 & 0.15 & 0.3086 & 0.0017 & 316.193 & 0.011 & 0.108 & 231.339 & 6.145 & 1 & 1 \\
4357037 & $f_{2a}$ & 346.22018 & 0.00056 & 2888.3354 & 0.0047 & 5.66 & 0.15 & 0.3463 & 0.0043 & 346.220 & 0.013 & 0.108 & 35.069 & \nodata & 1 & $-$1 \\
4357037 & $f_{2b}$ & 345.3365 & 0.0030 & 2895.726 & 0.025 & 1.06 & 0.15 & 0.451 & 0.023 & 345.338 & 0.012 & 0.098 & 1.359 & 7.391 & 1 & 0 \\
4357037 & $f_{2c}$ & 344.45564 & 0.00064 & 2903.1314 & 0.0054 & 4.93 & 0.15 & 0.3503 & 0.0049 & 344.456 & 0.013 & 0.110 & 26.544 & 7.405 & 1 & 1 \\
4357037 & $f_3$ & 549.7182 & 0.0036 & 1819.114 & 0.012 & 2.23 & 0.15 & 0.559 & 0.011 & 549.718 & 0.032 & 0.107 & 5.434 & \nodata & ? & ? \\
4357037 & $f_4$ & \nodata & \nodata & \nodata & \nodata & \nodata & \nodata & \nodata & \nodata & 956.11 & 0.29 & 0.318 & 3.284 & \nodata & ? & ? \\
\enddata
\end{deluxetable}
\clearpage
\end{landscape}

\end{document}